\documentclass[symmetry,article,submit,moreauthors,pdftex]{Definitions/mdpi}

\firstpage{1}
\pubvolume{1}
\issuenum{1}
\articlenumber{0}
\pubyear{2021}
\copyrightyear{2021}
\datereceived{}
\dateaccepted{}
\datepublished{}
\hreflink{https://doi.org/} 

\Title{The Primordial Particle Accelerator of the Cosmos}

\TitleCitation{Primordial Particle Accelerator of the Cosmos}

\Author{Asher Yahalom  $^{1,2}$\orcidA{}*}

\AuthorNames{Asher Yahalom}

\AuthorCitation{Yahalom, A.}

\address{%
$^{1}$ \quad Ariel University, Faulty of Engineering, Department of Electrical \& Electronic Engineering, Ariel 40700, Israel; asya@ariel.ac.il\\
$^{2}$ \quad Ariel University, Center for Astrophysics, Geophysics, and Space Sciences (AGASS),
Ariel 40700, Israel;}

\corres{Correspondence: asya@ariel.ac.il; Tel. +972-54-7740294: }

\abstract{In a previous paper we have shown that superluminal particles are allowed by
the general relativistic theory of gravity provided that the metric is locally Euclidean.
Here we calculate the probability density function of a canonical ensemble of  superluminal particles as function of temperature. This is done for both space-times invariant under
Lorentz symmetry group, and for space times invariant under an Euclidean symmetry group.
 Although only the Lorentzian metric is stable for normal matter density, an Euclidian metric can be created under special gravitational circumstances and persist in a limited region of space-time consisting of the very early universe which is characterized by extremely high densities and temperatures. Superluminal particles also allow attaining thermodynamic equilibrium at a shorter duration and also suggest a rapid expansion of the matter density, thus making mechanism such as inflation (which demands invoking and ad-hoc scalar field) redundant. This is in accordance with Occam's razor.}


\begin{document}
\newcommand{\beq} {\begin{equation}}
\newcommand{\enq} {\end{equation}}
\newcommand{\ber} {\begin {eqnarray}}
\newcommand{\enr} {\end {eqnarray}}
\newcommand{\eq} {equation}
\newcommand{\eqs} {equations }
\newcommand{\mn}  {{\mu \nu}}
\newcommand{\abp}  {{\alpha \beta}}
\newcommand{\ab}  {{\alpha \beta}}
\newcommand{\sn}  {{\sigma \nu}}
\newcommand{\rhm}  {{\rho \mu}}
\newcommand{\sr}  {{\sigma \rho}}
\newcommand{\bh}  {{\bar h}}
\newcommand{\br}  {{\bar r}}
\newcommand {\er}[1] {equation (\ref{#1}) }
\newcommand {\ern}[1] {equation (\ref{#1})}
\newcommand {\Ern}[1] {Equation (\ref{#1})}
\newcommand{\hdz}  {\frac{1}{2} \Delta z}
\newcommand{\curl}[1]{\vec{\nabla} \times #1} 
\newcommand {\Sc} {Schr\"{o}dinger}
\newcommand {\SE} {Schr\"{o}dinger equation }
\newcommand{\ce}  {continuity equation }

\section{Introduction}

It is well known that our daily space-time is approximately of
Lorentz (Minkowski) type with a metric $\eta_{\mn} = \ {\rm diag }
\ (1,-1,-1,-1)$ (we adopt the standard notation in which Greek indices take the values $0,1,2,3$)
and thus physical equations should be invariant under the Lorentz symmetry group. The above statement is taken as one of the
central assumptions of the theory of special relativity and has
been supported by numerous experiments. But one should ask why
should it be so?

Many textbooks \cite{Weinberg} state that in the general theory of relativity
any space-time is locally of the type $\eta_{\mn} = \ {\rm diag }
\ (1,-1,-1,-1)$, although it can not be presented so globally due
to the effect of matter. This is a part of the demands dictated by
the well known equivalence principle. The above principle is taken
to be one of the assumptions of general relativity other
assumption such as diffeomorphism invariance, and the requirement
that theory reduce to Newtonian gravity in the proper regime lead
to the Einstein equations:
\beq
G_\mn = -\frac{8 \pi G}{c^4} T_\mn
\label{ein}
\enq
in which $G_\mn$ is the Einstein tensor, $T_\mn$
is the stress-energy tensor, $G$ is the gravitational constant and $c$ is the velocity of light.

The Principle of Equivalence rests on the equality of gravitational and inertial mass,
demonstrated by Galileo, Huygens, Newton, Bessel, and E$\ddot{o}$tv$\ddot{o}$s. Einstein reflected that,
as a consequence, no external static homogeneous gravitational field could be detected
in a freely falling elevator, for the observers, their test bodies, and the elevator
itself would respond to the field with the same acceleration \cite{Weinberg}. This means that the observer will experience
himself as free, not feeling the effect of any force at all. Mathematically speaking for the observer
space time is locally (but not globally) flat and Minkowskian.

The point is that one need not assume that space-time is locally Lorentz
based on an empirical (unexplained) facts, rather one can derive this property from the field equations
based on the stability of the Minkowskian solution. Other unstable flat solutions of non Minkowskian type,
 such as an Euclidian metric $\eta_{\mn} = \ {\rm diag } \ (1,1,1,1)$ can exist in a limited region of space-time, in which the physical equations will have an Euclidean symmetry group.
 In an Euclidian metric there are no speed limitations and thus the alleged particle can travel in faster than light speed \cite{JMP}.
The reader should notice that already Eddington \cite[page 25]{Edd} has considered the possibility that the
universe contains different domains in which some domains are locally
Lorentzian and others have some other local metric of the type
 $\eta_{\mn} = \ {\rm diag } \ (-1,-1,-1,-1)$ or the type $\eta_{\mn} = \ {\rm diag } \ (1,1,-1,-1)$. The stability of those domains was not discussed by Eddington.

Many authors have suggested explanations to the locally Lorentzian nature of space-time
 \cite{Greensite4,Greensite6,Itin,vanDam}. What is common to all the above approaches is that
 additional theoretical structures \& assumptions are needed.
In previous works \cite{Yahaloma,Yahalomb,Yahalomc,Yahalome} it was shown that General relativistic equations and
 linear stability analysis suffice to obtain a unique choice of the Lorentzian metric being the only one which is stable.
Other metrics are allowed but are unstable and thus can exist in only a limited region of space-time.
The analysis will not be repeated here, the reader is referred to the original literature.
It should be mentioned that the choice of coordinates in the Fisher approach to physics is also
justified using the stability approach \cite{Yahalomd}. The nonlinear stability question of the
Lorentzian metric was settled by D. Christodoulou \&  S. Klainerman \cite{Nonlin}.
As for the nonlinear instability of other spaces of constant metric this remains an open
question at this time.

For non empty space-time the situation can be drastically
different \cite{Yahalomf}. The existence of the intuitive partition of 4
dimensional space into "spatial" space and "temporal" time, is a
feature of an almost empty space-time. This does not contradict
the fact that such a partition can not be demonstrated in general
solutions of Einstein's equations, such as the one discovered by G\"{o}del
\cite{Yourgrau}. But this problem is not a characteristic of
exotic space-times rather it is a property of standard
cosmological models.

Standard Cosmology has many fundamental problems those include the
horizon problem, the flatness problem, the entropy problem and the
monopole problem \cite{Narlikar}. A possible solution to those
problems were suggested by Alan Guth using his famous inflation
theory \cite{Guth}. Entropy problems which plagued the original
inflation model has led to a new inflation model suggested by
Linde \cite{Linde} which solve the entropy problem but required
fine tuning of parameters. The same criticism holds for chaotic
inflation also suggest by Linde \cite{Lindechaos}. On 17 March
2014, astrophysicists of the BICEP2 collaboration announced the
detection of inflationary gravitational waves in the B-mode power
spectrum, which if confirmed, would provide clear experimental
evidence for the theory of inflation \cite{BICEP2}. However, on 19
June 2014, lowered confidence in confirming the findings was
reported.

It is the opinion of the author of this paper that a basic flaw in
common to all inflation models. All inflation models require to
postulate one or more scalar fields which have no function,
implication or purpose in nature except for their ad-hoc use in
the inflation model. This is in sharp contradiction with the
principle of Occam's razor which demand that a minimum number of
assumptions will explain a maximum number of phenomena.
Postulating a physical field for every phenomena does not serve
the purpose of theoretical science. In the words of Einstein:
"Everything should be made as simple as possible, but not simpler". Moreover, it will be shown
that a perfectly good explanation within the frame-work of
standard Cosmology does exist for the horizon problem if one looks
closely at the metric changes of the Friedman-Lemaitre-Robertson-Walker metric.

The plan of this paper is as follows: in the first section we
describe possible mechanisms of metric change. In the following
section we describe a particle trajectory in a general flat space. Then
we analyze particle trajectories in Lorentz space-time for the standard subluminal cases.
The following section will discuss dynamics in the presence of an
Euclidean metric. Next we analyze particle trajectories in a Lorentz space-time but now we assume that the particles are superluminal. The final section is devoted to statistical analysis of free particles of the three different types  (Euclidian, subluminal Lorentzian and superluminal Lorentzian), in which we shall attempt to describe an equilibrium probability density function of a canonical ensemble of free particles. Then the possible physical implications
of the current theory are described. Finally some concluding remarks are given.

\section{Possible Mechanisms of Metric Change}

It was shown in \cite{Yahaloma} that among the possible flat space metrics only
the Lorentzian metric is stable and can persist for a considerable region of space-time.
Nevertheless one may still inquire if a mechanism exists by which a metric change does occur (in the sense that the eigen-values of the metric change signs), can we create some how a metric of the type $g_{\mn} = \ {\rm diag } \ (+1,+1,-1,-1)$
in some region of space-time? The answer obviously has to do with the only reason a metric should
 change according to equation (\ref{ein}) and this is $T_\mn$.
Looking at available solution of general relativity one finds that metric changes are quite common.

 The Schwarzschild square interval (in terms of spherical coordinates $t,r,\theta,\phi$) is given by:
\beq
c^2 d\tau^2 =
(1-\frac{r_s}{r})c^2 dt^2-\frac{dr^2}{1-\frac{r_s}{r}}-r^2(d\theta^2+\sin^2\theta
d\phi^2)
\label{Schwarzs}
\enq
In which $\tau$ is the proper time, and $r_s$ is the Schwarzschild radius (in meters) of the massive
body,
 which is related to its mass $M$ by $r_s = \frac{2GM}{c^2}$. It is obvious that while for $r>r_s$  the
 metric is locally (up to scaling) $g_{\mn} = \ {\rm diag } \ (+1,-1,-1,-1)$. For $r<r_s$  the
 metric is locally (up to scaling) $g_{\mn} = \ {\rm diag } \ (-1,+1,-1,-1)$. Hence the direction
 of temporal and (one) spatial axis is exchanged. Notice, however, that although the sign of the eigen-values did change we are still left with a Lorentzian metric.

 Another example is the Friedman-Lemaitre-Robertson-Walker square interval which is well known in cosmological models:
\beq
c^2 d\tau^2 = c^2 dt^2-a(t)^2 \left( \frac{dr^2}{1-\kappa r^2}+r^2(d\theta^2+\sin^2\theta d\phi^2)\right)
\label{Friedman}
\enq
$a(t)$ is known as the "scale factor" and $\kappa$ may be
taken to have units of length$^{-2}$, in which case $r$ has units
of length and a(t) is unitless. $\kappa$ is then the Gaussian
 curvature of the space at the time when $a(t) = 1$. Hence for radial distances such that $r<\frac{1}{\sqrt{\kappa}}$  the  metric is locally (up to scaling) $g_{\mn} = \ {\rm diag } \ (+1,-1,-1,-1)$ that is Lorentzian. However, for
$r>\frac{1}{\sqrt{\kappa}}$ the  metric is locally (up to scaling)
$g_{\mn} = \ {\rm diag } \ (+1,+1,-1,-1)$. This means that a
particle propagating in a radial direction will experience an
Euclidean metric.

 One should notice that in the above cases a signature change is accompanied by a metric singularity \cite{Donald}  while the signature changes considered by Eddington \cite{Edd} involve zeros. However, metric singularities are  not curvature singularities and can be removed by proper choice of coordinates.

It will be also interesting to find a metric which is completely
Euclidean in some regime of space-time, while being Lorentzian in another
 such a transitory metric may take the form
 \beq
 g_\mn =  \ {\rm diag } \ (+1,2e^{-\frac{(x_\mu-x_{0\mu})^2}{\Delta^2}}-1,2e^{-\frac{(x_\mu-x_{0\mu})^2}{\Delta^2}}-1,
 2e^{-\frac{(x_\mu-x_{0\mu})^2}{\Delta^2}}-1)
  \enq
  which is necessary to create an Euclidean domain of a width $\Delta$
 located at $x_{0\mu}$. More analytical effort is needed in order to describe accurately the conditions under which space-time  will become locally completely Euclidean. Here we shall simply assume that a similar  solution is attained at least in a limit region of space-time.

\section{Particle Trajectories in Flat Space}

Let us now look at a particle travelling in a space-time with a constant metric of arbitrary form.
Such a particle can be described by the Action ${\cal A}$:
\beq
{\cal A} = - m c \int  d \tau - e  \int A^\alpha d x_\alpha
\label{Action}
\enq
In the above $\tau$ is the trajectory interval:
\beq
d \tau^2 = \left|\eta^{\alpha \beta} d x_\alpha d x_\beta \right|
= \left| d x_\alpha d x^\alpha \right|
\label{length}
\enq
 $x_\alpha$  are the particle coordinates (raising and lowering indices is done using the metric as is customary), $m$ is the particle mass, $e$ is the particle charge and $A^\alpha$
are some functions of the particle coordinates (that transform as a four dimensional vector). Basic variational analysis leads to the following equations of motion:
\beq
m \frac{d u^\alpha}{d \tau}= -\frac{e}{c} u^\beta (\partial_\beta A^\alpha - \partial^\alpha A_\beta), \qquad u^\alpha \equiv \frac{d x^\alpha}{d \tau}
\label{Eqmot}
\enq
in which the metric $\eta_{\alpha \beta}$ can be of any flat type: Lorentzian, Euclidean etc.

\subsection{Lorentz Space-Time}

Let us assume a Lorentz Space-Time with a metric $\eta_\mn = \ {\rm diag } \ (1,-1,-1,\--1)$. Hence space-time
is dissected into spatial and temporal coordinates. The spatial coordinates are $\vec x = (x_1,x_2,x_3)$ and the
temporal coordinate is $x_0$. Since it is customary to measure time in different units (seconds) than space (meters)
we write $x_0 = c t$, in which $c$ serves as a units conversion factor. We now define the velocity: $\vec v \equiv \frac{d \vec x}{d t},
\quad v = |\vec v|$.
In a similar way we dissect $A_\alpha$ into temporal and spatial parts:
\beq
A_\alpha=(A_0,A_1,A_2,A_3) \equiv  (A_0, \vec A) \equiv (\frac{\phi}{c}, \vec A)
\label{Av}
\enq
the factor $\frac{1}{c}$ in the last term allows us to obtain the equations in MKS units, it
is not needed in other types of unit systems.
Using \ern{Av}, we can define a magnetic field:
\beq
\vec B = \vec \nabla \times \vec A
\label{magnetic}
\enq
($\vec \nabla$ has the standard definition of vector analysis) and an electric field:
\beq
\vec E =-\frac{\partial \vec A}{\partial t} -\vec \nabla \phi
\label{electric}
\enq
For subluminal particles $v<c$ we can than write $d \tau^2$ as:
\beq
d \tau^2 = c^2 dt^2 (1-\frac{v^2}{c^2}), \qquad d \tau = c dt \sqrt{1-\frac{v^2}{c^2}}
\label{lengthlsl}
\enq
And using the above equations one can write the spatial part of \ern{Eqmot} as:
\beq
 \frac{d }{d t}\left(m \frac{\vec v}{\sqrt{1-\frac{v^2}{c^2}}}\right)= e \left(\vec E + \vec v \times \vec B\right)
\label{Eqmotsl}
\enq
The above equation shows clearly that a subluminal particle in a Lorentz space must remain subluminal. Since as the
particle is accelerated to $c$ its "effective mass" $m_{eff} \equiv  \frac{m}{\sqrt{1-\frac{v^2}{c^2}}}$ becomes infinite.
On the other hand for superluminal particles (which are $v>c$ at $\tau=0$) we can  write $d \tau^2$ as:
\beq
d \tau^2 = c^2 dt^2 (\frac{v^2}{c^2}-1), \qquad d \tau = c dt \sqrt{\frac{v^2}{c^2}-1}
\label{lengthlfl}
\enq
And using the above equations one can write the spatial part of \ern{Eqmot} as:
\beq
 \frac{d }{d t}\left(m \frac{\vec v}{\sqrt{\frac{v^2}{c^2}-1}}\right)= e \left(\vec E + \vec v \times \vec B\right)
\label{Eqmotfl}
\enq
Here the difficulty would be near the velocity $c$, in which its "effective mass" $m_{eff} \equiv  \frac{m}{\sqrt{\frac{v^2}{c^2}-1}}$ becomes infinite.
In the absence of forces the velocity of the above
particle remains constant and superluminal. We conclude that in a Lorentz space time there is
a difficulty to pass the velocity $c$ from below as is well known. In any case luminal particle
with $v=c$ have $d \tau =0$ which make this parameter unsuitable to describe the trajectory for those type of particles.

\subsection{Euclidean Space-Time}

Let us assume an Euclidean space-time with a metric $\eta_\mn = \ {\rm diag } \ (+1,+1,\-+1,+1)$. Here space-time
is dissected (arbitrarily) into spatial and temporal coordinates as in the Lorentz space which are
measured in the customary units.  Again we dissect $A_\alpha$ into temporal and spatial parts as in \ern{Av}. Using \ern{Av}, we can define the magnetic field as in \ern{magnetic} but the electric field is defined now as:
\beq
\vec E =-\frac{\partial \vec A}{\partial t} +\vec \nabla \phi
\label{electrice}
\enq
notice that this definition for the electric field is different than in the Lorentz space but is necessary in order to maintain Faraday's law.
For all particles either (subluminal or superluminal) we can than write $d \tau^2$ as:
\beq
d \tau^2 = c^2 dt^2 (1+\frac{v^2}{c^2}), \qquad d \tau = c dt \sqrt{1+\frac{v^2}{c^2}}
\label{lengthle}
\enq
And using the above equations one can write the spatial part of \ern{Eqmot} as:
\beq
 \frac{d }{d t}\left(m \frac{\vec v}{\sqrt{1+\frac{v^2}{c^2}}}\right)= e \left(\vec E - \vec v \times \vec B\right)
\label{Eqmote}
\enq
The above equation shows clearly that particles in an Euclidean space are quite indifferent to passing the velocity $c$.

\section{Statistical Physics}

The definition of a probability density function is intimately connected to the notion of phase space. This in turn arises naturally when time is the independent variational variable. The path to the Hamiltonian formalism goes through defining a Lagrangian at the action per unit time and through
the Lagrangian one can define the canonical momenta and finally the Hamiltonian. We shall follow this route.

\subsection{Lagrangian and Canonical Momenta}

\begin{figure}
\centering
\includegraphics[width= 0.7 \columnwidth]{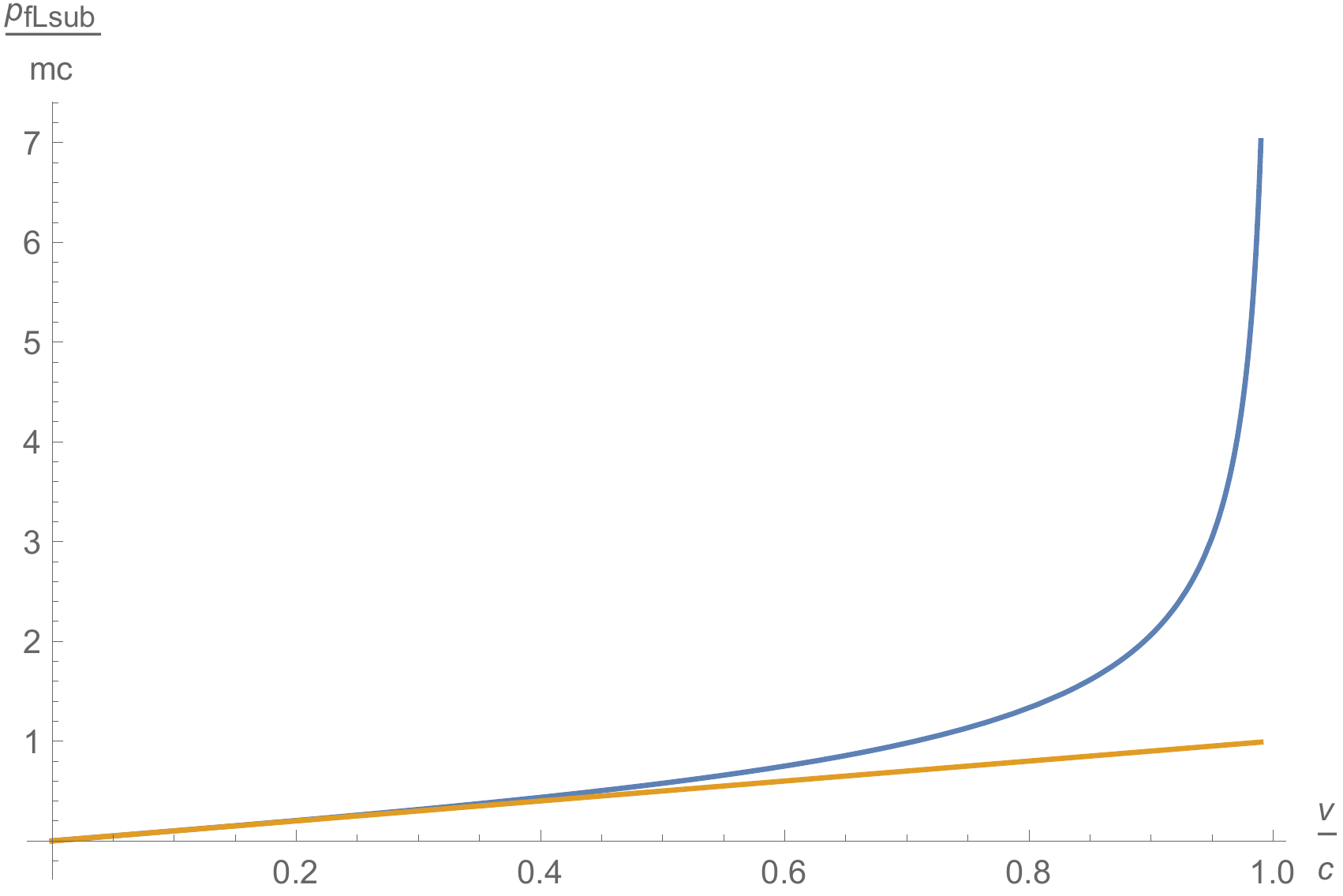}
 \caption{Free momentum for Lorentzian subluminal particles, the correct expression is compared with the classical one. The blue line is the correct expression while the orange line is the classical approximation.}
 \label{pfLsubf}
\end{figure}

\begin{figure}
\centering
\includegraphics[width= 0.7 \columnwidth]{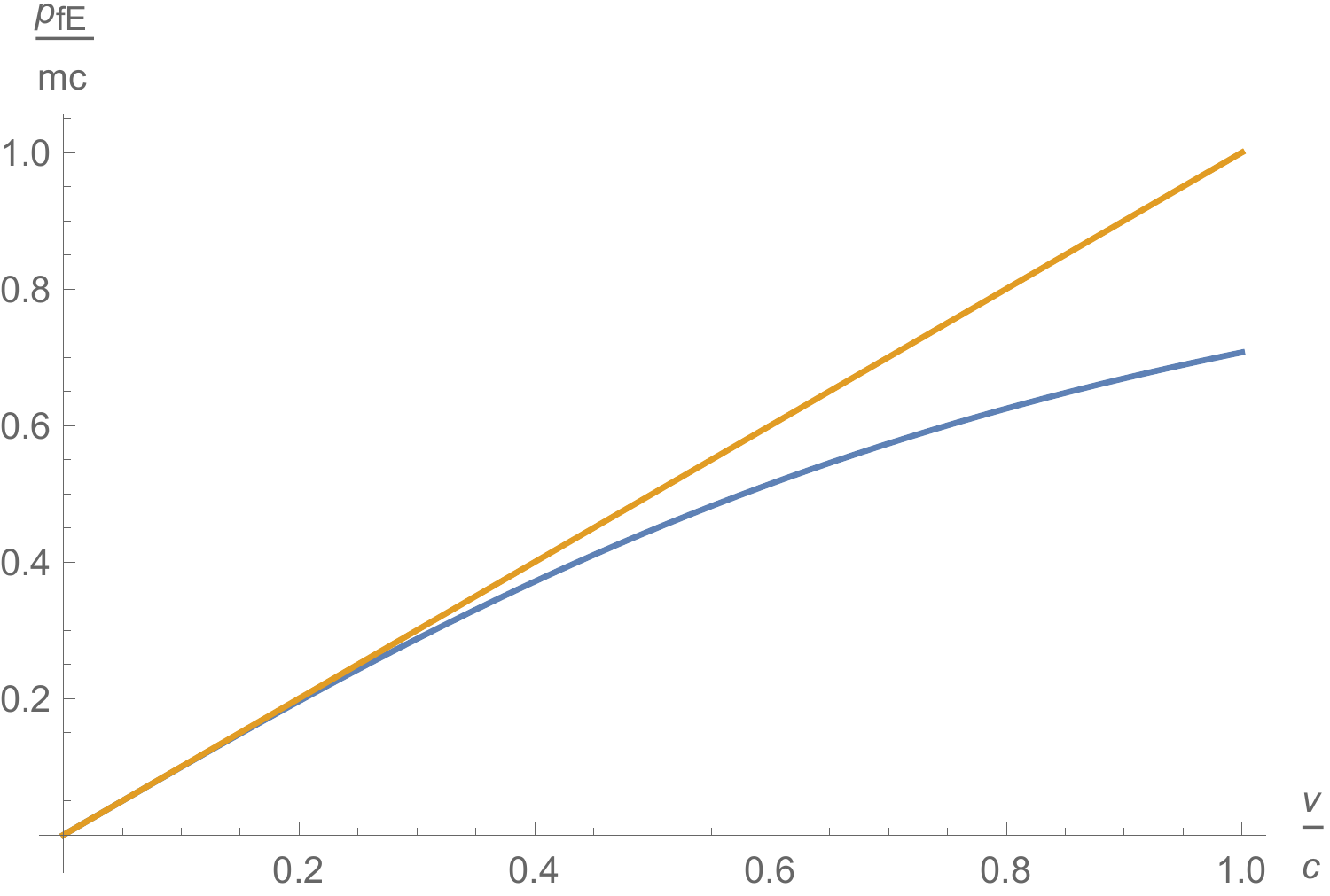}
 \caption{Free momentum for Euclidean subluminal particles, the correct expression is compared with the classical one. The blue line is the correct expression while the orange line is the classical approximation.}
 \label{pfEf0}
\end{figure}

\begin{figure}
\centering
\includegraphics[width= 0.7 \columnwidth]{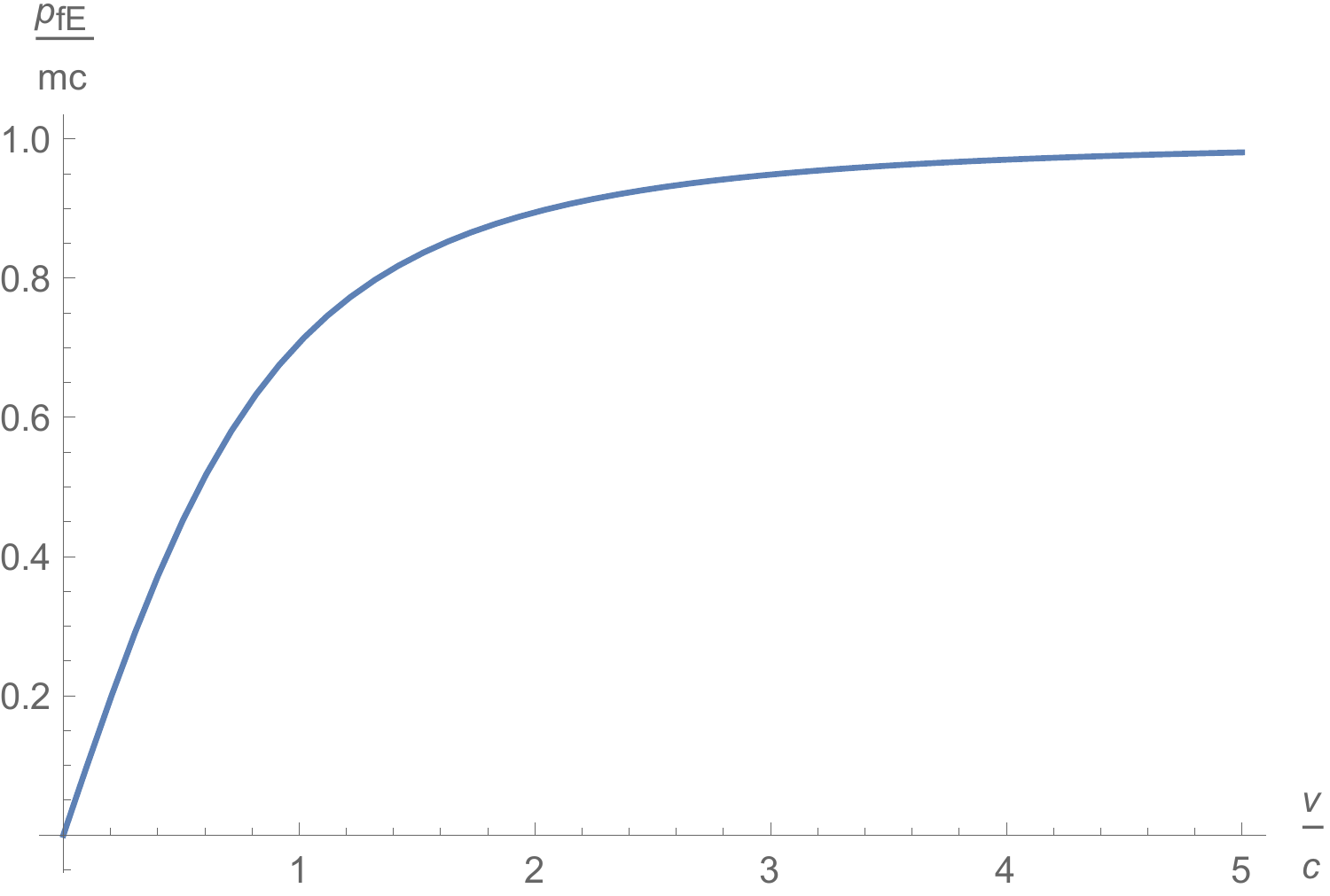}
 \caption{Free momentum for Euclidean particles.}
 \label{pfEf}
\end{figure}

\begin{figure}
\centering
\includegraphics[width= 0.7 \columnwidth]{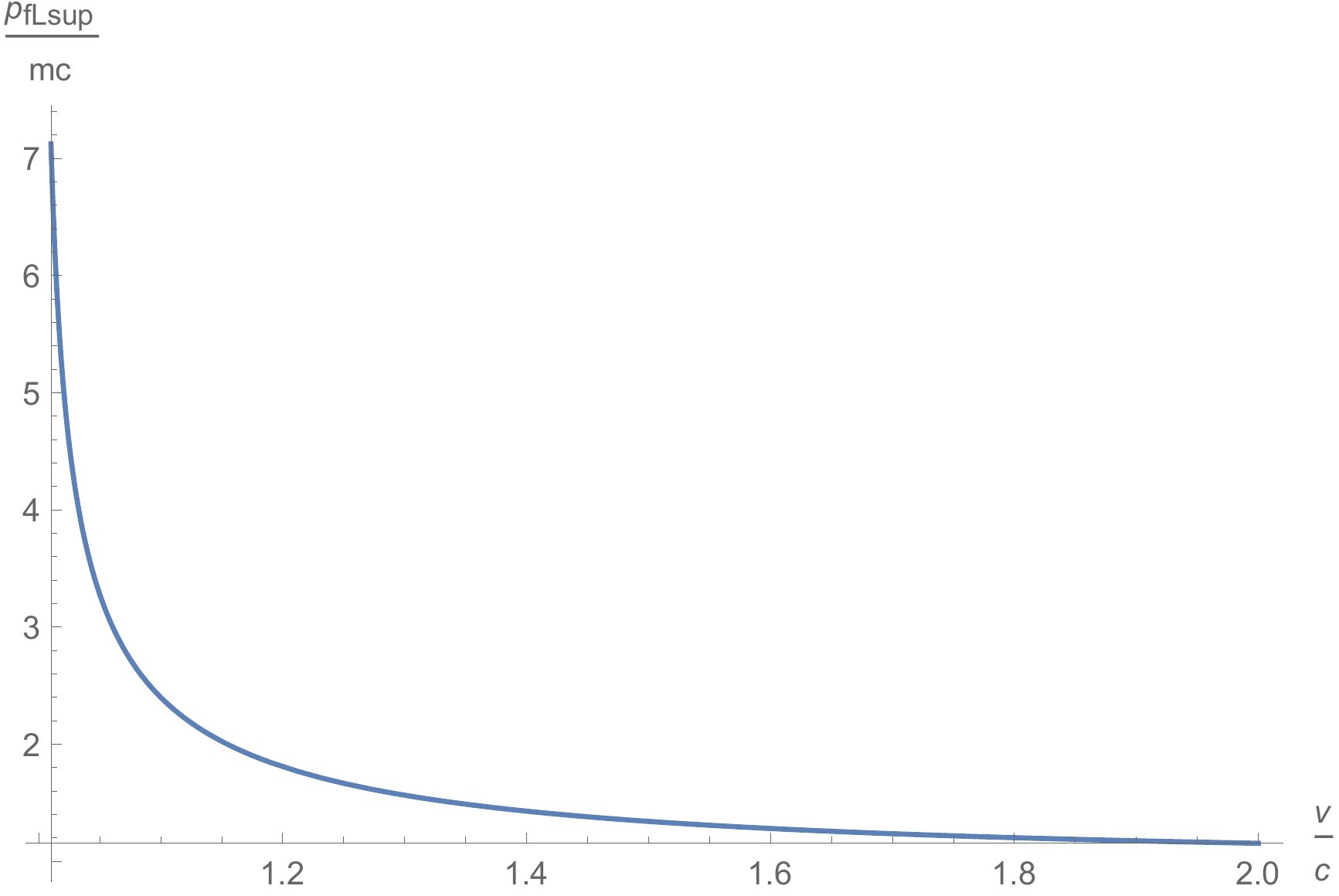}
 \caption{Free momentum for Lorentzian superluminal particles.}
 \label{pfLsupf}
\end{figure}

\begin{figure}
\centering
\includegraphics[width= \columnwidth]{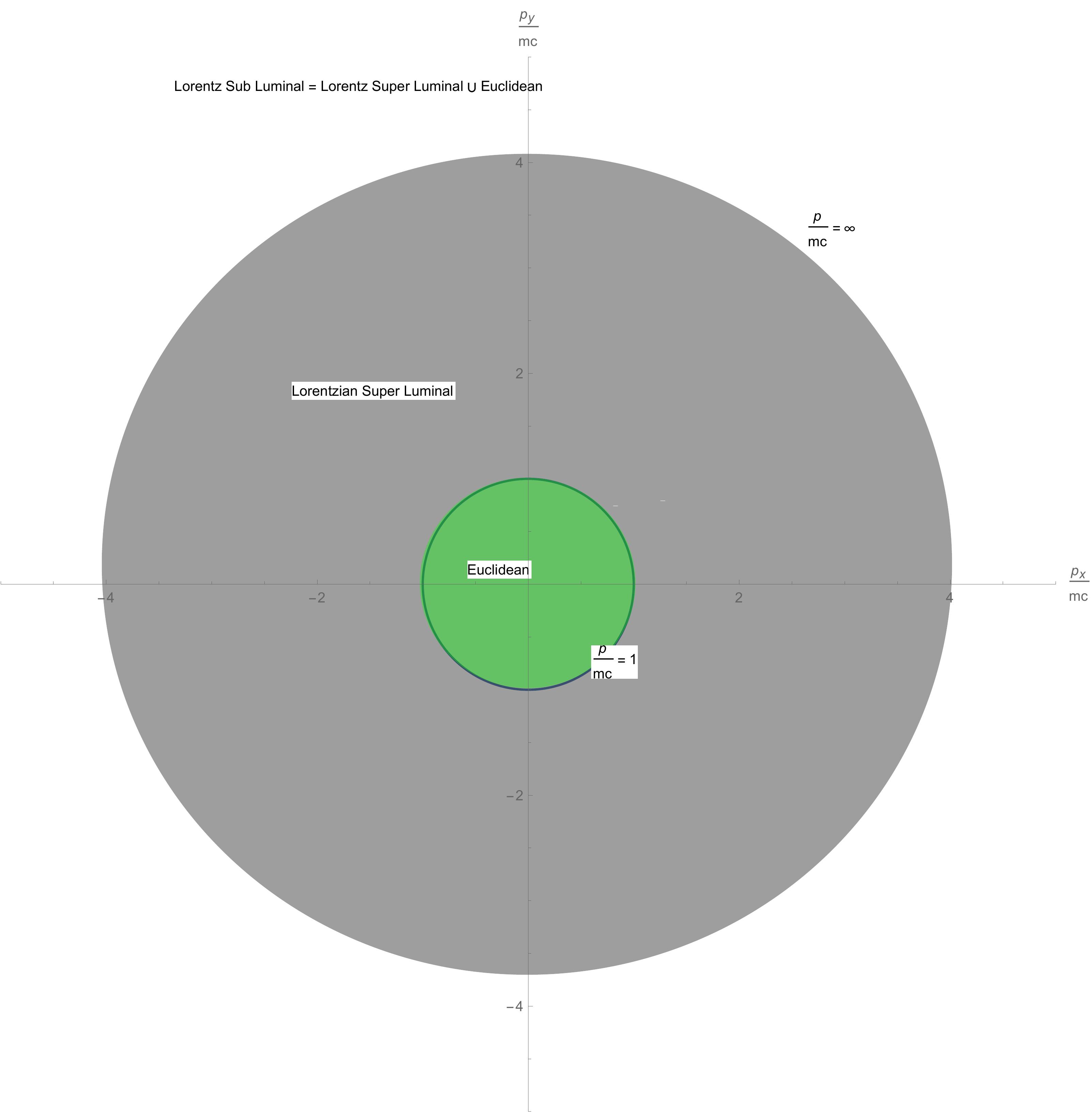}
 \caption{A depiction of the cross section of the momentum space for the cases considered: Euclidean (a compact sphere) Lorentzian superluminal (an infinite domain with a spherical hall) and Lorentzian subluminal (the entire momentum plane). The later is a union of the former.}
 \label{phasespacef}
\end{figure}

Let us write the action given in \ern{Action} as:
\beq
{\cal A} = - m c \int  d \tau - e  \int A^\alpha d x_\alpha
= - \int dt \left[ m c  \frac{d \tau}{d t}+ e A^\alpha \frac{d x_\alpha}{d t} \right]
\label{Action2}
\enq
Introducing the notation:
\beq
v_\alpha \equiv \frac{d x_\alpha}{d t} = (c,\vec v)
\label{val}
\enq
and using the definition of \ern{length} it follows that:
\beq
\frac{d \tau}{d t} = \sqrt{\left| \frac{d x_\alpha}{d t} \frac{d x^\alpha}{d t} \right|}
= \sqrt{\left| v_\alpha  v^\alpha  \right|} = \sqrt{\left| c^2 + v_i  v^i  \right|}
\label{dtau}
\enq
in which the Latin indices are $ i \in \{1,2,3\}$. The upper index $v^i$ has the following meaning:
\beq
v^i = \left\{ \begin{array}{cc}
                 + v_i & {\rm Euclidean~ metric} \\
                 - v_i & {\rm Lorentz~ metric}
               \end{array}
               \right.
\label{vi}
\enq
Thus:
\ber
\frac{d \tau}{d t} &=& \left\{ \begin{array}{cc}
                 \sqrt{\left| c^2 +\vec v^2  \right|} & {\rm Euclidean~ metric} \\
                 \sqrt{\left| c^2 -\vec v^2  \right|} & {\rm Lorentz~ metric}
               \end{array}
               \right.
 \nonumber \\
               &=& \left\{ \begin{array}{cc}
                           \sqrt{c^2 + v^2} & {\rm Euclidean~ metric} \\
                           \sqrt{c^2 -  v^2} & {\rm Lorentz~ metric~subluminal~case}\quad v<c \\
                           \sqrt{v^2 -  c^2} & {\rm Lorentz~ metric~superluminal~case}\quad v>c
                         \end{array},
               \right.
\label{dtau2}
\enr
It follows from \ern{Action2} that one can define a Lagrangian:
\ber
{\cal A} = \int dt L, \qquad
L &\equiv& - \left[ m c  \frac{d \tau}{d t}+ e A^\alpha \frac{d x_\alpha}{d t} \right]
 = - \left[ m c  \sqrt{\left| v_\alpha  v^\alpha  \right|}+ e A^\alpha v_\alpha \right]
 \nonumber \\
  &=& - \left[ m c  \sqrt{\left| c^2 + v_i  v^i  \right|} + e c A^0 + e A^i v_i \right].
\label{Action3}
\enr
This leads to a canonical momentum of the form:
\beq
p^i \equiv \frac{\partial L}{\partial v_i} = -mc \frac{\pm v^i}{\sqrt{\left| c^2 + v_i  v^i  \right|}} - e A^i
\label{momentumdef}
\enq
the sign is decided according whether the absolute value changes or does not change the sign of  $c^2 + v_i  v^i$. It does not change sign of course in the standard subluminal Lorentzian case but also the Euclidean case. On the other the superluminal Lorentzian case involves a sign change. Hence:
\beq
p^i  = \left\{\begin{array}{cc}
                 -mc \frac{v^i}{\sqrt{\left| c^2 + v_i  v^i  \right|}} - e A^i & {\rm Lorentzian~subluminal~or~ Euclidean} \\
                 +mc \frac{v^i}{\sqrt{\left| c^2 + v_i  v^i  \right|}} - e A^i & {\rm Lorentzian~superluminal}
              \end{array}\right.
\label{momentum2}
\enq
Introducing the standard notation:
\beq
\beta_i \equiv \frac{v_i}{c}, \quad \beta^i \equiv \frac{v^i}{c},
\quad \beta \equiv \frac{v}{c},\qquad
\gamma \equiv \frac{1}{\sqrt{\left| 1 + \frac{v_i  v^i}{c^2}  \right|}}
= \frac{1}{\sqrt{\left| 1 + \beta_i \beta^i   \right|}}
\label{stannot}
\enq
in which:
\beq
\gamma = \left\{ \begin{array}{cc}
                          \frac{1}{\sqrt{ 1 + \frac{v^2}{c^2}}}  & {\rm Euclidean~ metric} \\
                           \frac{1}{\sqrt{1 - \frac{v^2}{c^2}}} & {\rm Lorentz~ metric~subluminal~case}\quad v<c \\
                           \frac{1}{\sqrt{\frac{v^2}{c^2}-1}} & {\rm Lorentz~ metric~superluminal~case}\quad v>c
                         \end{array}
               \right. .
\label{gamcases}
\enq
We may write:
\beq
p^i  = \left\{\begin{array}{cc}
                 - \gamma m  v^i - e A^i & {\rm Lorentzian~subluminal~or~ Euclidean} \\
                 + \gamma m  v^i - e A^i & {\rm Lorentzian~superluminal}
              \end{array}\right.
\label{momentum3}
\enq
We can define a "free" momentum as:
\beq
p_f^i  = p^i + e A^i = \left\{\begin{array}{cc}
                 - \gamma m  v^i  & {\rm Lorentzian~subluminal~or~ Euclidean} \\
                 + \gamma m  v^i  & {\rm Lorentzian~superluminal}
              \end{array}\right.
\label{momentumf3}
\enq
this is a misnomer, as the case $A^i = 0$ includes the free particle case but applies also to the case in which the particle moves under the influence of a scalar electric potential. Defining a canonical momentum vector we have:
\beq
\vec p \equiv (p^1,p^2,p^3)  = \left\{
\begin{array}{cc}
  \gamma m  \vec v + e  \vec A &  {\rm Lorentzian~subluminal} \\
    -\gamma m  \vec v - e  \vec A & {\rm Euclidean} \\
      -\gamma m  \vec v + e  \vec A & {\rm Lorentzian~superluminal}
       \end{array}
                                \right.
\label{momentum4}
\enq
in which we have used \ern{vi} and also:
\beq
A^i = \left\{ \begin{array}{cc}
                 + A_i & {\rm Euclidean~ metric} \\
                 - A_i & {\rm Lorentz~ metric}.
               \end{array}
               \right.
\label{Ai}
\enq
And similarly:
\beq
\vec p_f \equiv (p_f^1,p_f^2,p_f^3) = \left\{
\begin{array}{cc}
  \gamma m  \vec v &  {\rm Lorentzian~subluminal} \\
    -\gamma m  \vec v  & {\rm Euclidean} \\
      -\gamma m  \vec v & {\rm Lorentzian~superluminal}.
       \end{array}
                                \right.
\label{momentum4f}
\enq
Interestingly the "free" momentum has the same direction as the velocity only in the (standard) Lorentzian subluminal case, in all other cases the momentum direction is the opposite. The
Lorentzian subluminal case reduces to the classical result for low velocities:
\beq
v \ll c \Rightarrow \gamma \simeq 1 \Rightarrow \vec p  \simeq m  \vec v + e  \vec A
\label{momentum5}
\enq
In all cases the magnitude of the "free" momentum is:
\beq
p_f = |\vec p_f| = \gamma m  v.
\label{pf}
\enq
Thus a subluminal Lorentzian "free" particle will have zero momentum for $v=0$ and infinite momentum for $v=c$ in which the momentum
is an increasing function of $v$ (see figure \ref{pfLsubf}), hence the phase space is non compact.
In the Euclidean case the small velocity momentum is similar to the Lorentzian case, however,
the momentum space in this case is limited inside a "momentum sphere" hence it is compact.
Thus a subluminal Lorentzian "free" particle will have zero momentum for $v=0$ and a momentum
of $p_{fE}=mc$ for $v=\infty$. The momentum is an increasing function of velocity which is bounded.
(see figures \ref{pfEf0} and \ref{pfEf}).
Finally for a superluminal Lorentzian particle we have $p_{fLsup}=\infty$ for the case $v=c$ and
$p_{fLsup}=mc$ for the case $v=\infty$. The momentum is a decreasing function of velocity
(see figure \ref{pfLsupf}) which differs considerably from our habits and physical intuition.
The momentum space is not compact, however, it has spherical hole in its middle of radius $mc$. One could say that the set union of the momentum space in the Euclidean case and the momentum space in the superluminal Lorentzian case is equal to the momentum space of the subluminal Lorentzian case, this situation is depicted in figure \ref{phasespacef}.
In terms of the $\gamma$ notation we can write the Lagrangian as:
\beq
L = - \left[\frac{m c^2}{\gamma} + e c A^0 + e A^i v_i \right].
\label{Lagrang}
\enq
For small velocities in a Lorentzian space time we have:
\beq
\gamma \simeq 1 + \frac{1}{2} \frac{v^2}{c^2}, \qquad \gamma^{-1} \simeq 1 - \frac{1}{2} \frac{v^2}{c^2}.
\label{gammas}
\enq
Hence the classical Lagrangian would be:
\beq
L_c = \frac{1}{2} m v^2 - (m c^2 + e \phi - e \vec A \cdot \vec v)
= \frac{1}{2} m v^2  + e \vec A \cdot \vec v - e \phi - m c^2.
\label{Lagrangc}
\enq
Finally we will be interested in the dependence of $v$ on $p_f$, first notice that:
\beq
m \gamma = \frac{p_f}{v} \Rightarrow m^2 v^2 = p_f^2 \gamma^{-2}
= p_f^2 \left| 1 + \frac{v_i  v^i}{c^2}  \right|.
\label{mgam}
\enq
Thus we obtain:
\beq
v = \left\{
\begin{array}{cc}
  \frac{p_f}{\sqrt{m^2+\frac{p_f^2}{c^2}}} &  {\rm Lorentzian~subluminal} \\
    \frac{p_f}{\sqrt{m^2-\frac{p_f^2}{c^2}}} & {\rm Euclidean} \\
      \frac{p_f}{\sqrt{\frac{p_f^2}{c^2}-m^2}} & {\rm Lorentzian~superluminal}
       \end{array}
                                \right.
\label{vofpf}
\enq

\subsection{Hamiltonian and Energy}

\begin{figure}
\centering
\includegraphics[width= 0.7 \columnwidth]{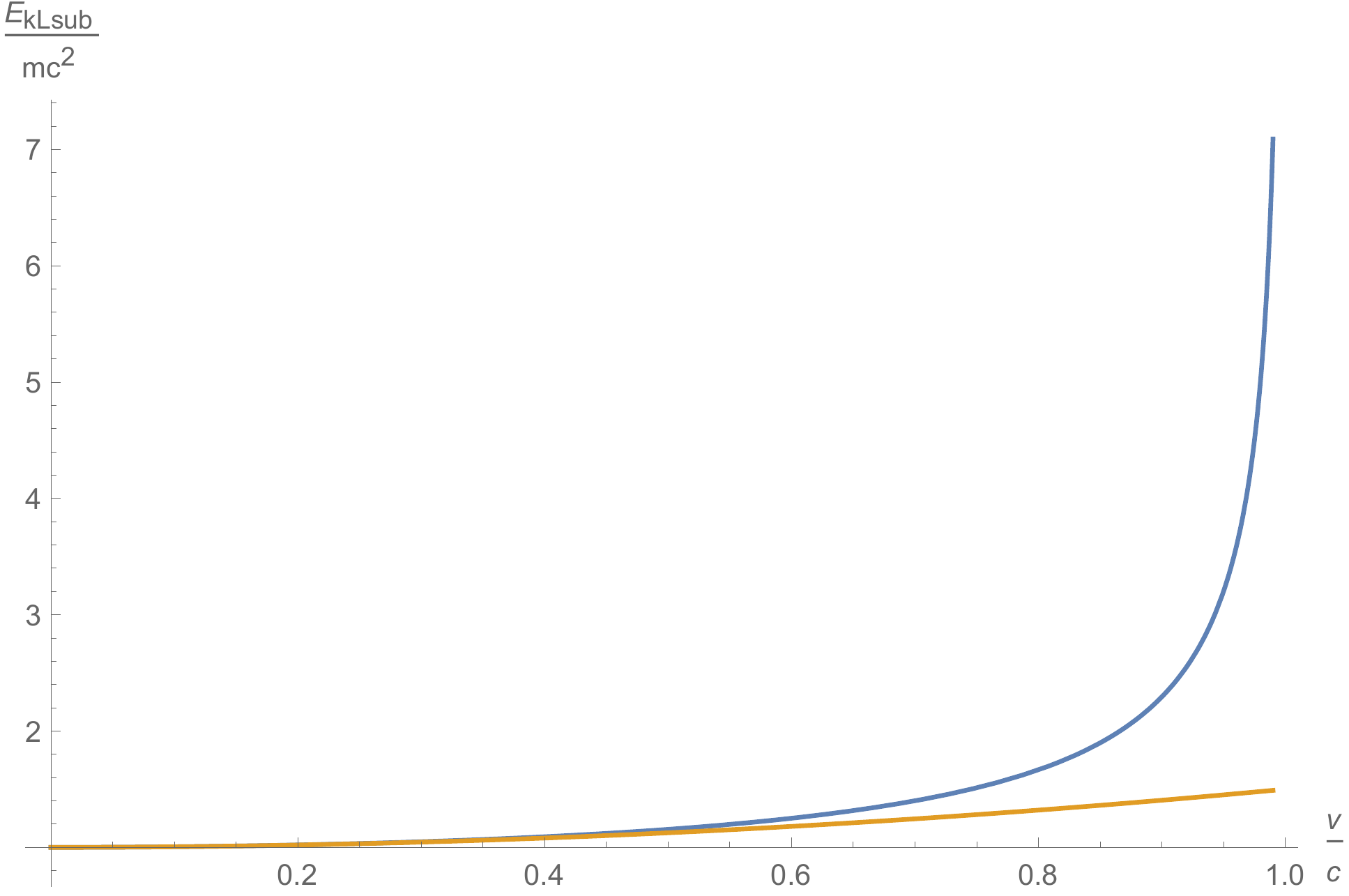}
 \caption{Kinetic energy for Lorentzian subluminal particles, the correct expression is compared with the classical one. The blue line is the correct expression while the orange line is the classical approximation}
 \label{EkLsubf}
\end{figure}

\begin{figure}
\centering
\includegraphics[width= 0.7 \columnwidth]{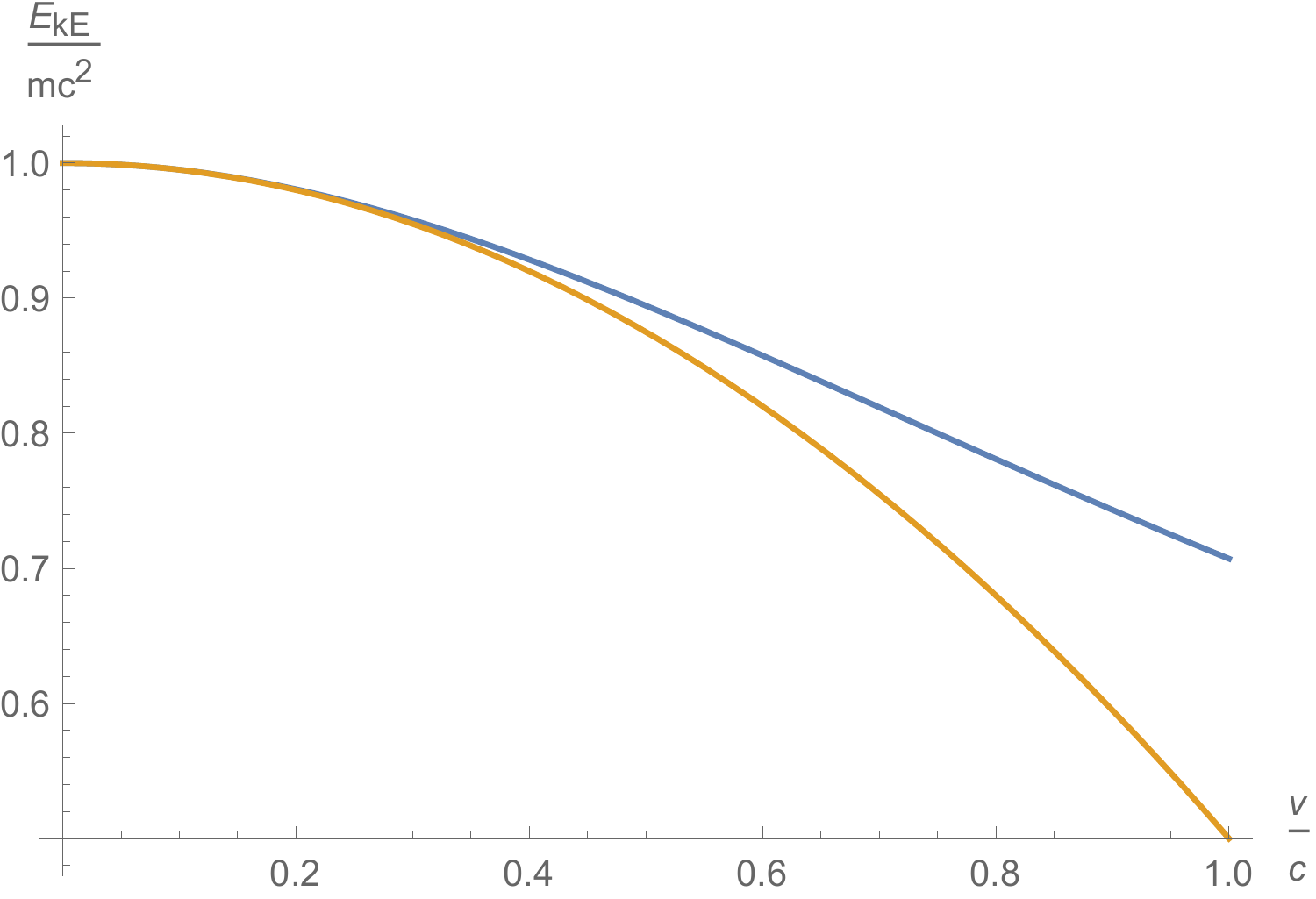}
 \caption{Kinetic energy for Euclidean particles, the correct expression (blue) is compared with the classical approximation (orange).}
 \label{EkEf}
\end{figure}

\begin{figure}
\centering
\includegraphics[width= 0.7 \columnwidth]{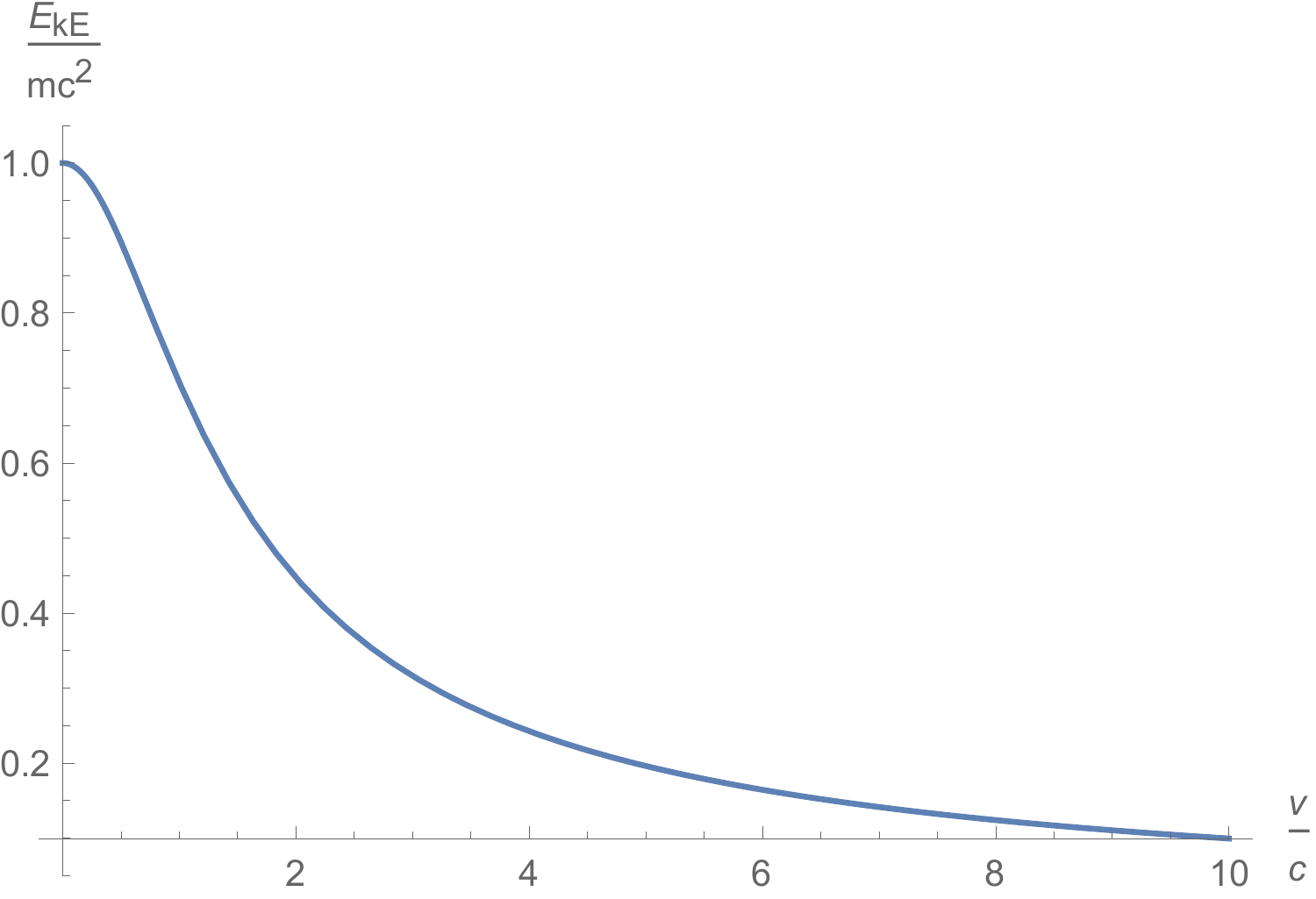}
 \caption{Kinetic energy for Euclidean particles of subluminal and superluminal velocities.}
 \label{EkEf2}
\end{figure}

\begin{figure}
\centering
\includegraphics[width= 0.7 \columnwidth]{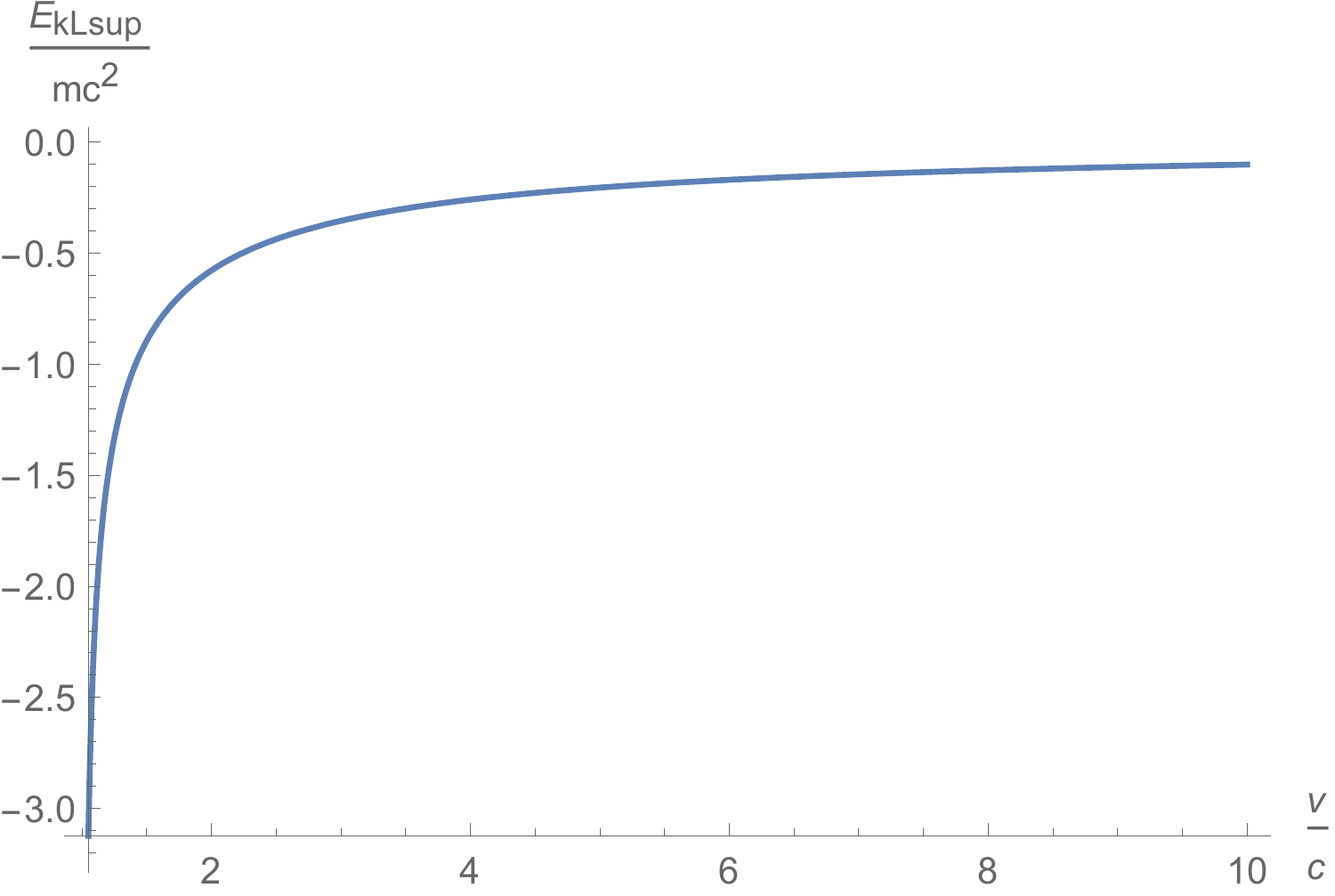}
 \caption{Kinetic energy for Lorentzian particles of superluminal velocities.}
 \label{EkLsupf}
\end{figure}

Once we have the canonical momenta and Lagrangian it is straight forward to calculate the Hamiltonian:
\ber
H &=& \vec v \cdot \vec p - L = p^i v_i - L = p^i v_i + \frac{m c^2}{\gamma} + e c A^0 + e A^i v_i
\nonumber \\
&=& (\mp \gamma m  v^i - e A^i) v_i+ \frac{m c^2}{\gamma} + e c A^0 + e A^i v_i
= \mp \gamma m  v^i v_i + \frac{m c^2}{\gamma} + e c A^0.
\label{Hamiltonian}
\enr
in the above the minus sign is for the Euclidean and subluminal Lorentzian cases while the plus sign
is for the superluminal Lorentzian case. This can be simplified as follows:
\ber
H &=& \gamma m (\mp v^i v_i + \frac{ c^2}{\gamma^2}) + e \phi =\gamma m \left(\mp v^i v_i + c^2 \left|1 + \beta_i \beta^i\right| \right)+ e \phi
\nonumber \\
&=& \gamma m \left( \mp v^i v_i +  \left|c^2 + v_i v^i \right| \right)+ e \phi
 = \pm \gamma m c^2 + e \phi.
\label{Hamiltonian2}
\enr
The plus sign in the last terms belongs to the Euclidean and subluminal Lorentzian cases while the minus sign is for the superluminal Lorentzian case. Explicitly:
\beq
H = \left\{  \begin{array}{cc}
               \gamma m c^2 + e \phi & {\rm Euclidean~ and ~subluminal~ Lorentzian} \\
               -\gamma m c^2 + e \phi & {\rm superluminal~Lorentzian}
             \end{array} \right.
\label{Hamiltonian2b}
\enq
The energy of the particle is $En=H$ and will remain constant as long as $\phi$ does not explicitly depend on time. The structure of the Hamiltonian is:
\beq
H = E_k + e \phi
\label{Hstr}
\enq
in which:
\beq
E_k \equiv \left\{ \begin{array}{cc}
                          \frac{m c^2}{\sqrt{ 1 + \frac{v^2}{c^2}}}  & {\rm Euclidean~ metric} \\
                           \frac{m c^2}{\sqrt{1 - \frac{v^2}{c^2}}} & {\rm Lorentz~ metric~subluminal~case}\quad v<c \\
                           -\frac{m c^2}{\sqrt{\frac{v^2}{c^2}-1}} & {\rm Lorentz~ metric~superluminal~case}\quad v>c
                         \end{array}
               \right. .
\label{Ek}
\enq
The kinetic energy has the following attributes. For the Lorentz subluminal case (which is the standard case) the minimal value for the kinetic energy is the rest energy $E_{kLsub~min} = m c^2$
obtained for $v=0$. However, it can reach an infinite value for velocities approaching the speed of light in vacuum $c$:
\beq
E_{kLsub~max} = \lim_{v \rightarrow c} E_{kLsub} = +\infty.
\label{EkLsmax}
\enq
It is always positive for all values of $v$. In the classical case in which $v \ll c$ we can partition the energy into a "classical kinetic energy" and a rest energy:
\beq
E_{kLsub} \simeq m c^2 + E_{kLsubc}, \qquad E_{kLsubc} \equiv \frac{1}{2} m v^2, \qquad v \ll c.
\label{EkLsubc}
\enq
The expression $E_{kLsub}$ is depicted in figure \ref{EkLsubf}.
In the Euclidean case the kinetic energy is always positive, it has a maximal value for a particle in rest: $E_{kE~max} = m c^2$ and a minimal value of zero for a particle travelling at an infinite speed, we recall that there are no speed limitation in an Euclidean space-time.
\beq
E_{kE~min} = \lim_{v \rightarrow +\infty} E_{kE} =  0.
\label{EkEmin}
\enq
Curiously one can define a "classical kinetic energy" also in the Euclidean case, but it will be negative:
\beq
E_{kE} \simeq m c^2 + E_{kEc}, \qquad E_{kEc} \equiv -\frac{1}{2} m v^2, \qquad v \ll c.
\label{EkEc}
\enq
The expression $E_{kE}$ is depicted in figure \ref{EkEf} for subluminal velocities
and in figure \ref{EkEf2} for superluminal velocities.
Finally we consider the superluminal Lorentzian kinetic energy, which differs from the previous cases in the attribute that it is always non positive. Its maximal value of zero is attained for infinite velocities and it can reach minus infinity when the velocity of the particle is reduced down to the speed of light $c$, a limit it cannot reach. Thus:
\beq
E_{kLsup~max} = \lim_{v \rightarrow +\infty} E_{kLsup} =  0, \qquad
E_{kLsup~min} = \lim_{v \rightarrow c} E_{kLsup} =  -\infty.
\label{EkLsup}
\enq
of course there is no sense of discussing the classical limit in this case, as by definition  superluminality requires $v>c$. The expression $E_{kLsup}$ is depicted in figure \ref{EkLsupf}.
Let us suppose that the potentials is time independent and the energy is conserved, it follows that
the energy $En$ is conserved and for any two points  $\vec x_1$ and $\vec x_2$ on the trajectory:
\beq
En = E_{k1} + e \phi_1 = E_{k2} + e \phi_2 \Rightarrow E_{k2} = E_{k1} + e (\phi_1 - \phi_2)
\label{encon}
\enq
Thus the kinetic energy can be increased or decreased using a potential difference, this is of course well known and is used for accelerating and decelerating charged particles in electrostatic accelerators like the tandem accelerator located in Ariel university \cite{Gover}. For a subluminal particle an increase in the kinetic energy means an increase in velocity and thus using a potential
difference a charged particle can be accelerated. Similarly by using a potential difference with an opposite sign the particle becomes slower as it kinetic energy is reduced. For Euclidean particles the situation is opposite, particles with higher kinetic energy are slower and with low kinetic energy are faster, nevertheless, potential differences can be still used to achieve acceleration and deceleration. Superluminal Lorentzian particles are similar to the subluminal Lorentzian particles in the sense that the (negative) kinetic energy would be lower for slower particles and higher for
faster particles. Another important difference between Euclidean and Lorentzian particles is lack of velocity limits in the former. In fact it is easy to see that using a finite energy equal to its rest mass ($m c^2$) a particle can be accelerated from zero velocity to infinite velocity in an Euclidean space-time. This is of course impossible for a Lorentzian particle. In the sub luminal case we will need an infinite amount of energy to accelerate the particle to the speed of light, while for a superluminal particle
an infinite amount of energy will be needed to reduce its speed to the speed of light. This entails an infinite potential difference and thus is physically impossible. We remark that even if the electromagnetic potentials are time dependent (as in the popular accelerator scheme of RF Linacs \cite{Balal}) this limitations cannot be avoided because that
although the total energy of the particle can be increased in this scenario and is not necessarily constant, it cannot increase to infinite values (which should be supplied from an infinite reservoir). Thus in a Lorentzian universe subluminal and superluminal particles must be separated by their velocities forever. We will discuss the cosmological and other implications of those facts later in this paper.

To conclude this subsection we would like to express the Hamiltonian as a function of the coordinates and the momenta. Using \ern{Hamiltonian2b} and equations (\ref{mgam}) and (\ref{vofpf}) this can be written as follows:
\beq
H = \left\{
\begin{array}{cc}
     c^2 \sqrt{m^2+\frac{p_f^2}{c^2}}+ e \phi &  {\rm Lorentzian~subluminal} \\
      c^2 \sqrt{m^2-\frac{p_f^2}{c^2}}+ e \phi & {\rm Euclidean} \\
      - c^2 \sqrt{\frac{p_f^2}{c^2}-m^2}+ e \phi & {\rm Lorentzian~superluminal}
       \end{array} \right. .
\label{Hp}
\enq
For Lorentzian classical particles \ern{momentum5} holds hence $p_f \simeq m v$  and thus:
\beq
H_c \simeq m c^2 + \frac{p_f^2}{2m} + e \phi
\label{cH}
\enq
Using \ern{momentumf3} we arrive at:
\beq
H (\vec x, \vec p) = \left\{
\begin{array}{cc}
     c^2 \sqrt{m^2+\frac{(\vec p - e \vec A (\vec x))^2}{c^2}}+ e \phi (\vec x) &  {\rm Lorentzian~subluminal} \\
      c^2 \sqrt{m^2-\frac{(\vec p + e \vec A (\vec x))^2}{c^2}}+ e \phi (\vec x) & {\rm Euclidean} \\
      - c^2 \sqrt{\frac{(\vec p - e \vec A (\vec x))^2}{c^2}-m^2}+ e \phi (\vec x) & {\rm Lorentzian~superluminal}
       \end{array}
                                \right.
\label{Hp2}
\enq
the first line in the above is Jackson's \cite{Jackson} formula (12.17). In the classical case
we have:
\beq
H_c (\vec x, \vec p) \simeq m c^2 + \frac{(\vec p - e \vec A (\vec x))^2}{2m} + e \phi (\vec x)
\label{cH2}
\enq

\subsection{Statistical Physics of "Classical" Particles}

Once we have a phase space we may discuss what form of the probability density function for a
particle to be in a specific part of this space. Elementary considerations show \cite{Bloch} that in thermal equilibrium this function must depend on the constants of motion of the system, in particular its energy.
\beq
f_{system} =  f_{system} (H) =  f_{system} (E_{system})
\label{fsys1}
\enq
Further if the system can be partitioned into two sub systems $A$ and $B$ of which the interaction is negligible (as in the case of free particles) it follows that:
\beq
 f_{system} (E_{system}) =  f_{system} (E_A + E_B) = f_A (E_A) f_B (E_B)
\label{fsys2}
\enq
This leads after a few trivial steps to the result that:
\beq
 f = \frac{e^{-\beta_T H}}{Z}
\label{fsys3}
\enq
$Z$ the normalization constant also known as the partition function. $\beta_T = \frac{1}{k_B T}$, in which $k_B$ is the Boltzmann constant:
\beq
k_B \equiv 1.380649 ~ 10^ {-23} {\rm ~m^2 ~kg ~s^{-2} ~K^{-1}}
\enq
and T is the temperature measured in degrees Kelvin. In what follows we will consider only free particles, more over we will consider only a single free particle. In this case:
\beq
 Z = \int e^{-\beta_T H} d^3 p
\label{fsinpart}
\enq
In all the relevant cases $H$ depends only the absolute value $p=|\vec p|$, hence the integral simplifies to:
\beq
 Z = 4 \pi \int e^{-\beta_T H} p^2 dp
\label{fsinpart2}
\enq
 We will start with the more familiar case of a subluminal Lorentzian particle and consider the more exotic cases later.

 \subsubsection{A Low speed Lorentzian particle}

 For a small velocity free Lorentzian particle \ern{cH2} takes the form:
 \beq
H_c = m c^2 + \frac{p^2}{2m}
\label{cH2f}
\enq
hence:
\beq
 f (\vec p) = \frac{e^{-\beta_T m c^2} e^{-\beta_T \frac{p^2}{2m}}}{Z}.
\label{fsyscf}
\enq
The partition function can be calculated to be:
\beq
 Z = 4 \pi \int e^{-\beta_T H_c } p^2 dp = (2 \pi m k_B T)^\frac{3}{2} e^{-\beta_T m c^2}
\label{Zcf}
\enq
Thus we obtain the well known Maxwell-Boltzmann probability density function:
\beq
 f (\vec p) = \frac{e^{- \frac{p^2}{2m k_B T}}}{(2 \pi m k_B T)^\frac{3}{2}}.
\label{fsyscf2}
\enq
This is a typical Gaussian distribution with a null average and a variance which is linear in the temperature and a standard deviation which is the square root of the same:
\beq
E[p^i]  = 0, \quad E[{p^i}^2] = m k_B T, \quad \sigma_{p^i} = \sqrt{m k_B T}.
\label{momentscf}
\enq
Using \ern{EkLsubc} we obtain the well known result for the average of the classical kinetic energy:
\beq
E[E_{kLsubc}]=E[\frac{1}{2} m v^2]=E[\frac{p^2}{2 m}] =\frac{3 m k_B T}{2 m} = \frac{3}{2} k_B T
\label{avergaenergycf}
\enq
Thus the average of the total kinetic energy which includes a rest energy term:
\beq
E[E_{kLsub}] \simeq m c^2 +  \frac{3}{2} k_B T.
\label{avergaenergycf2}
\enq
In what follows expressions will appear simpler using a normalized momenta and $\beta_T$ defined
as follows:
\beq
\vec p' \equiv \frac{\vec p}{m c}, \qquad \lambda \equiv \beta_T  m c^2 =\frac{m c^2}{k_B T}.
\label{normali}
\enq
Thus low $\lambda$ means high temperature, and high $\lambda$ means low temperature.
In terms of those we may write the classical distribution as:
\beq
 f (\vec p') = (\frac{\lambda}{2 \pi})^\frac{3}{2} e^{- \frac{\lambda p'^2}{2}}, \qquad
 Z' = (\frac{2 \pi}{\lambda})^\frac{3}{2} e^{-\lambda} .
\label{fsyscf3}
\enq
Thus:
\beq
 \lim_{\lambda \rightarrow 0} Z'  = \infty, \qquad  \lim_{\lambda \rightarrow \infty} Z'  = 0.
 \label{Zpcf}
\enq
In term of $\lambda$ the average energy becomes:
\beq
 \bar{E}'_{kLsub} = \frac{E[E_{kLsub}]}{m c^2} \simeq 1 + \frac{3}{2 \lambda}
\label{avergaenergycfp}
\enq
for high $\lambda$ (low velocities).

\subsubsection{A Lorentzian particle}

\begin{figure}
\centering
\includegraphics[width= 0.7 \columnwidth]{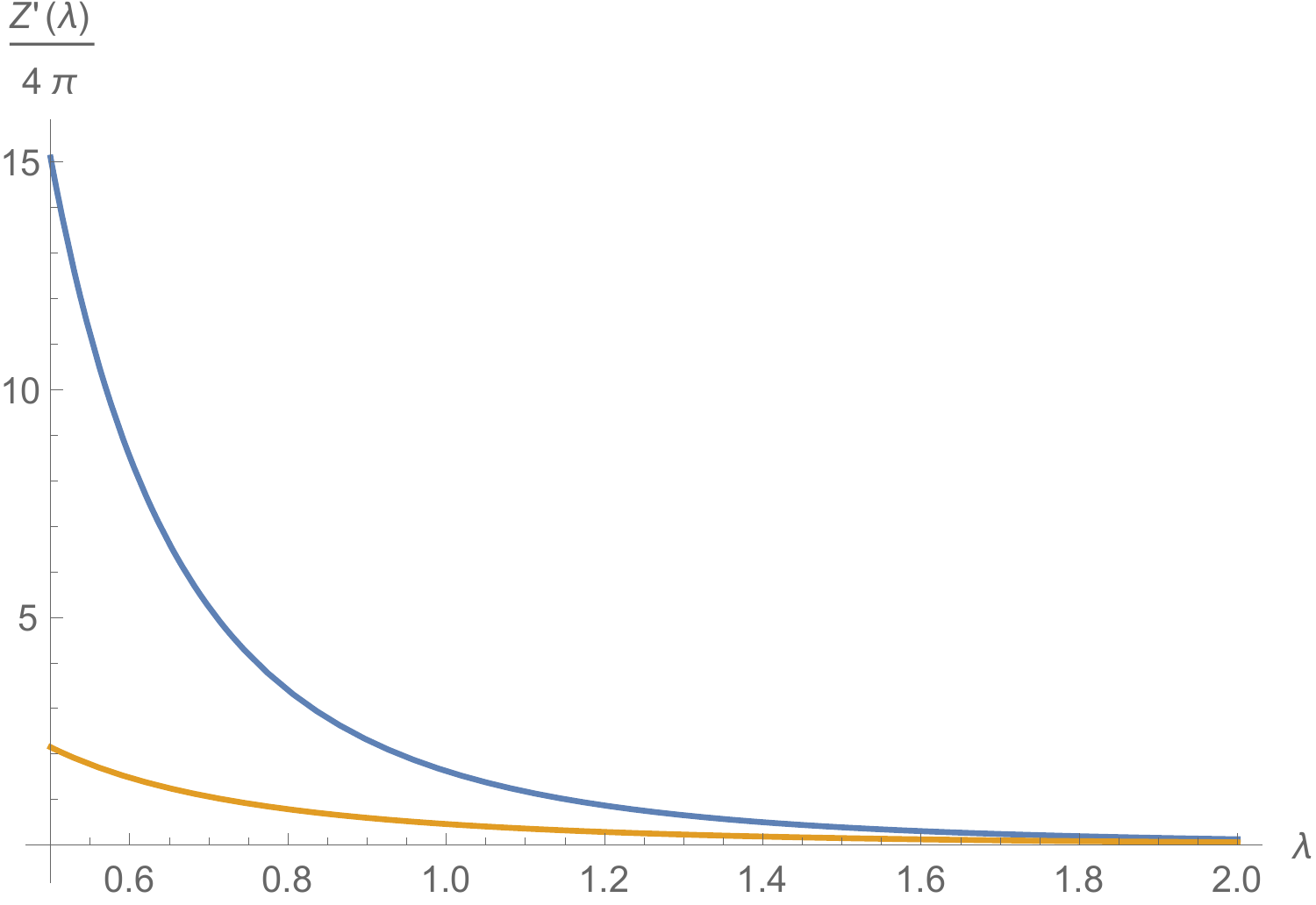}
 \caption{Partition function for a free subluminal Lorentzian particle, the blue line is the correct value while the orange represents the low velocity approximation.}
 \label{ZLsub}
\end{figure}

\begin{figure}
\centering
\includegraphics[width= 0.7 \columnwidth]{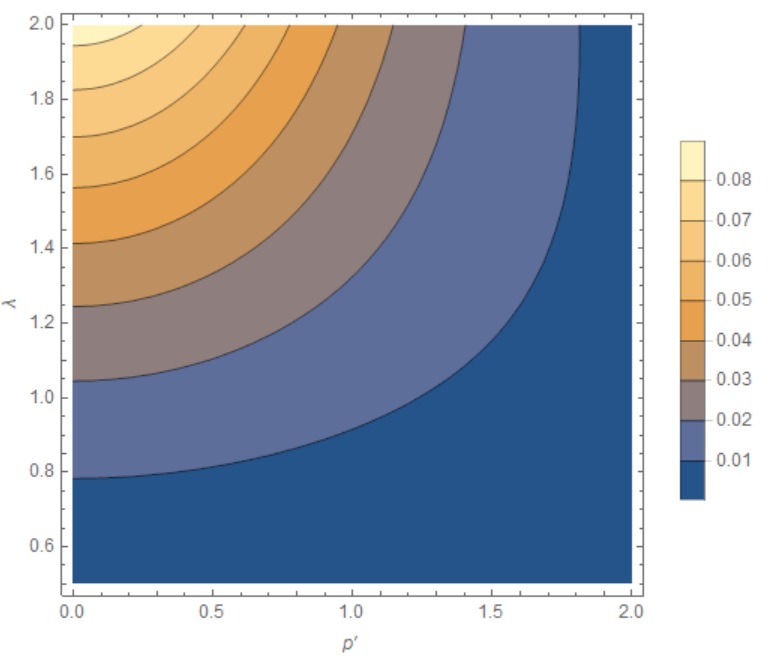}
 \caption{Probability density function for a free subluminal Lorentzian particle as function of $p'$ and $\lambda$.}
 \label{fLsub2D}
\end{figure}

\begin{figure}
\centering
\includegraphics[width= 0.7 \columnwidth]{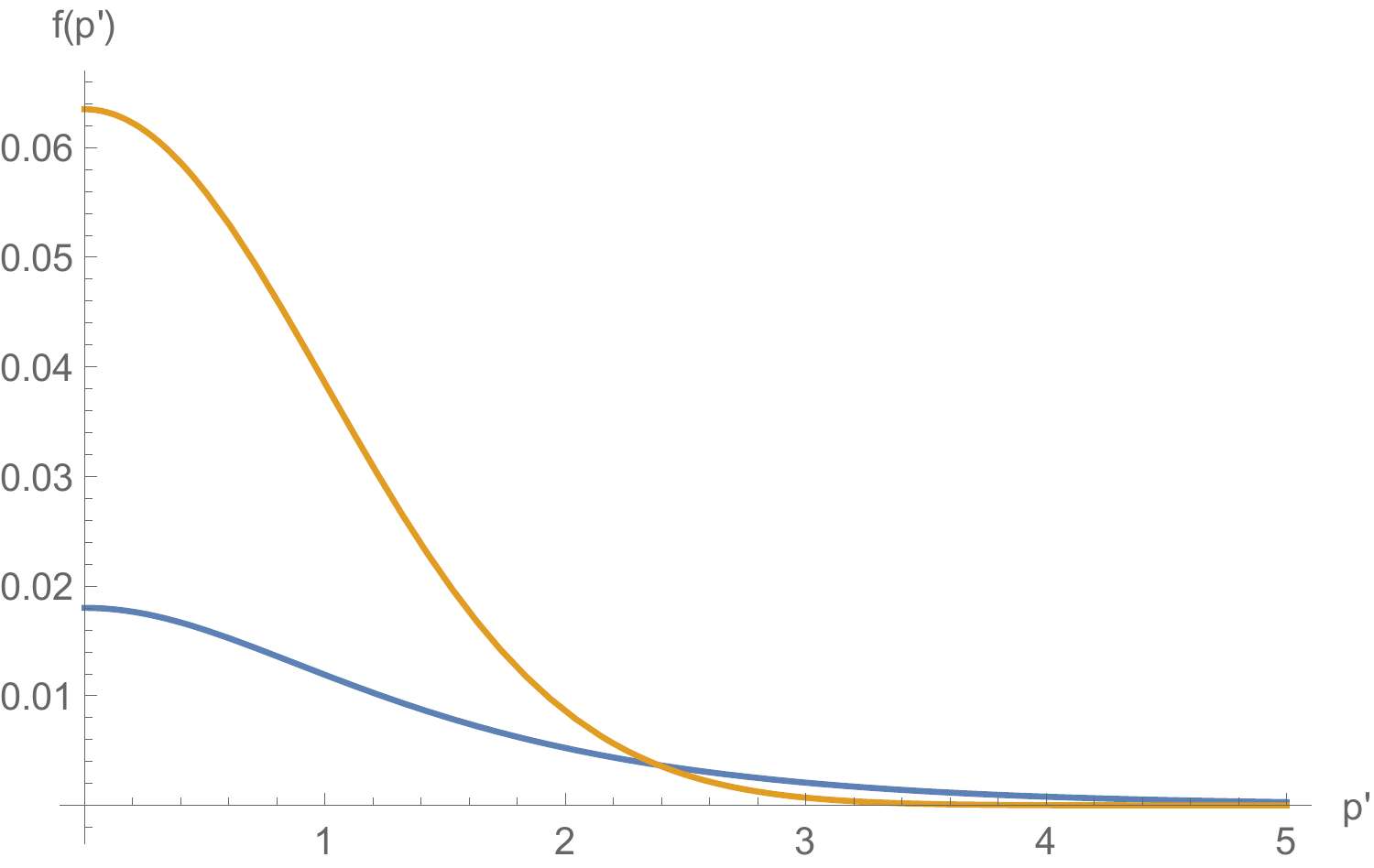}
 \caption{Probability density function for a free subluminal Lorentzian particle as function of $p'$ for $\lambda = 1$. The blue line depicts the correct value, while the orange line depicts the Maxwell-Boltzmann approximation.}
 \label{fLsubl}
\end{figure}

\begin{figure}
\centering
\includegraphics[width= 0.7 \columnwidth]{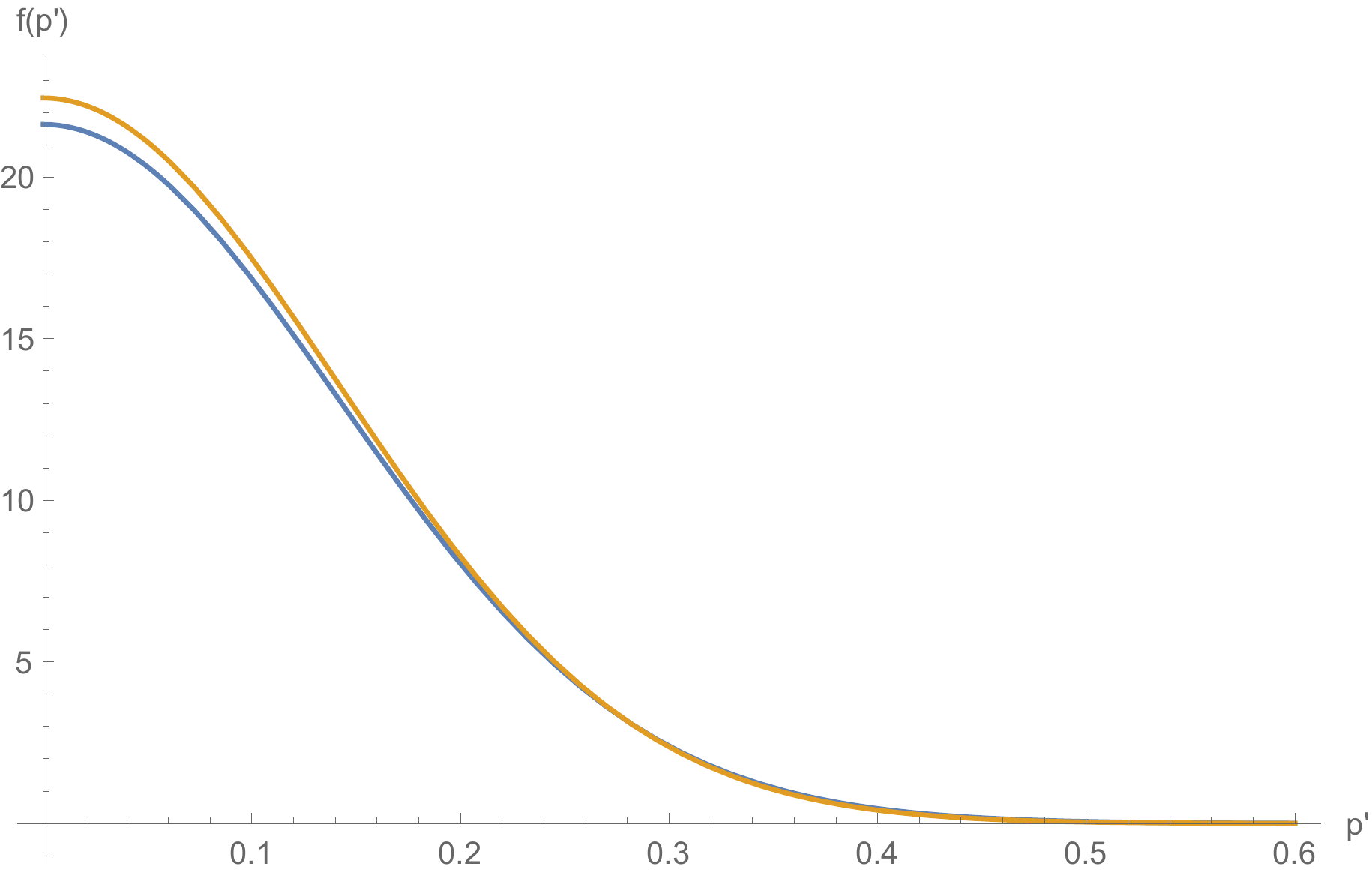}
 \caption{Probability density function for a free subluminal Lorentzian particle as function of $p'$ for $\lambda = 50$. The blue line depicts the correct value, while the orange line depicts the Maxwell-Boltzmann approximation.}
 \label{fLsubh}
\end{figure}

\begin{figure}
\centering
\includegraphics[width= 0.7 \columnwidth]{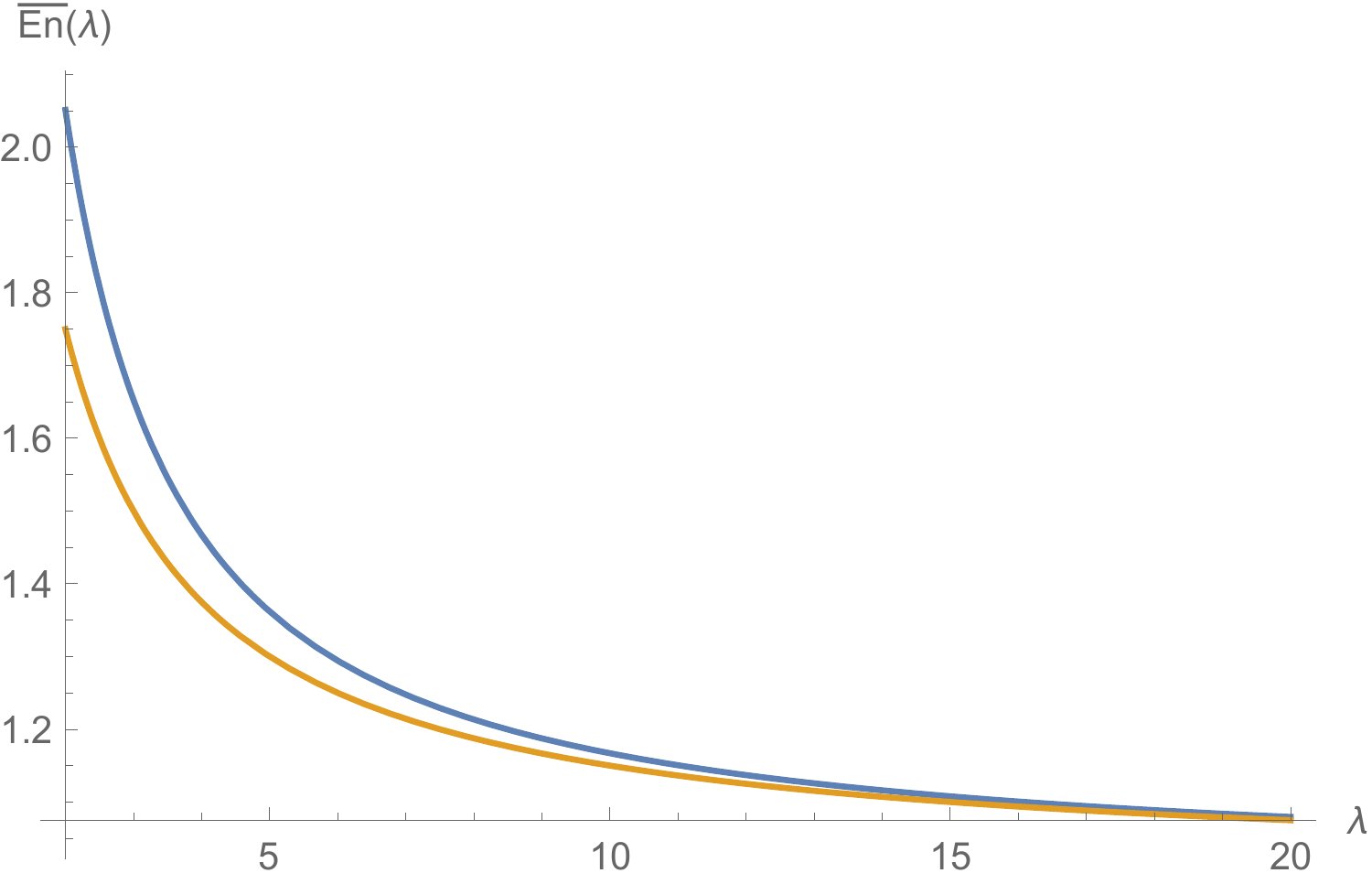}
 \caption{Average energy for free Lorentzian subluminal particles with high $\lambda$. The blue line depicts the correct value, while the orange line depicts the Maxwell-Boltzmann approximation.}
 \label{Eavh}
\end{figure}

\begin{figure}
\centering
\includegraphics[width= 0.7 \columnwidth]{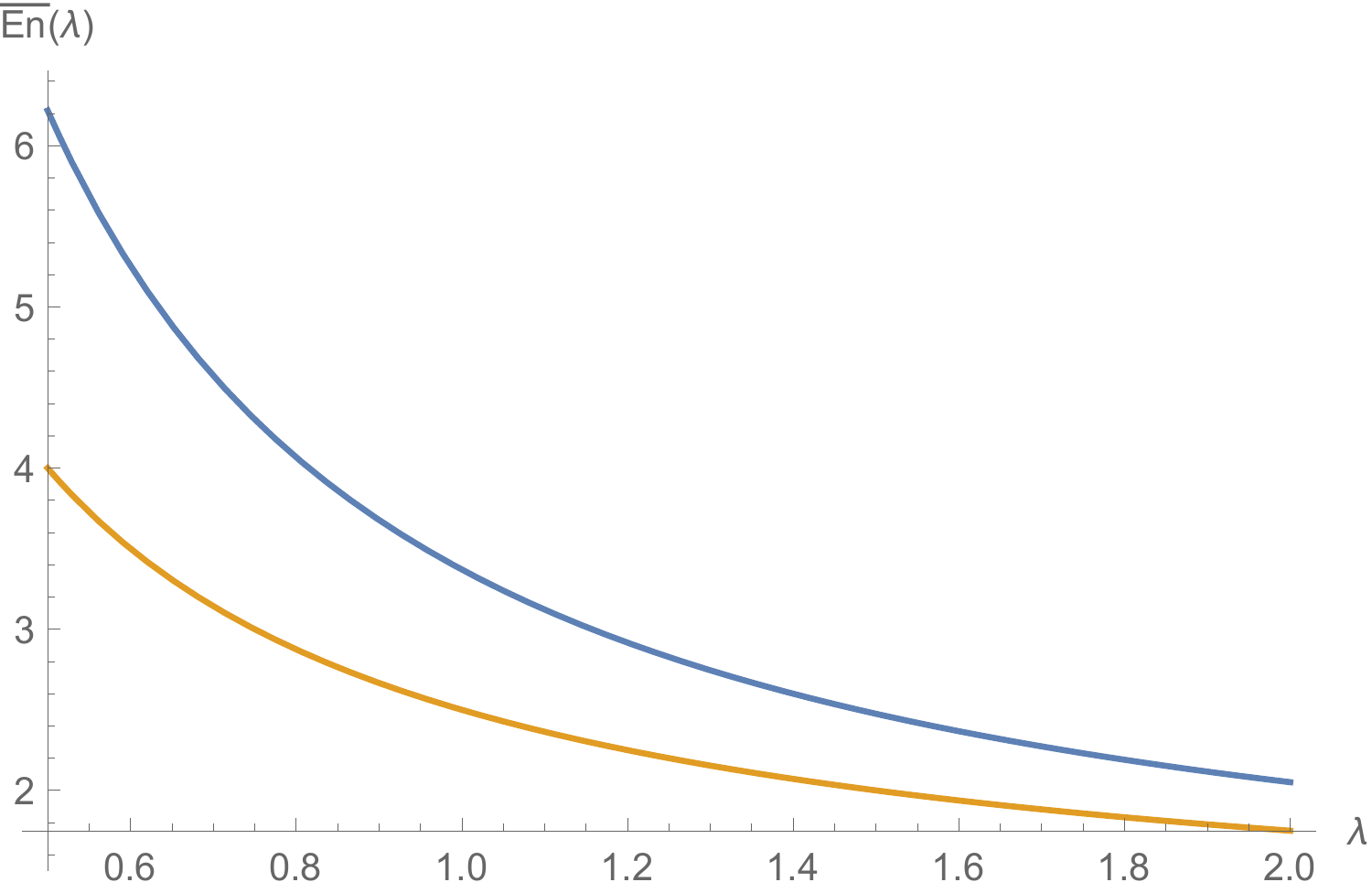}
 \caption{Average energy for free Lorentzian subluminal particles with high $\lambda$. The blue line depicts the correct value, while the orange line depicts the Maxwell-Boltzmann approximation.}
 \label{Eavl}
\end{figure}

Generally speaking the Maxwell-Boltzmann probability density function does not describe the momentum
distribution function for a Lorentzian particle unless the velocities are much smaller than the speed of light. Taking into account \ern{Hp2} for a free particle, and \ern{fsys3} we arrive at:
\beq
 f (\vec p') = \frac{e^{-\beta_T H}}{Z'} = \frac{e^{-\lambda \sqrt{1+p'^2}}}{Z'}
\label{fLsub}
\enq
in which:
\beq
 Z' (\lambda) =  4 \pi \int_{0}^{\infty} e^{-\lambda \sqrt{1+p'^2}} p'^2 dp'
\label{ZpLsub}
\enq
The above expression cannot be evaluated analytically, but can be easily evaluated numerically
(see figure \ref{ZLsub}) in which we compare the results to the classical case. As can be clearly seen the results converge for high $\lambda$ (low temperature) but differ considerably for small $\lambda$ (high temperature).
For a Lorentzian subluminal particle we have:
\beq
 \lim_{\lambda \rightarrow 0} Z'  = \infty, \qquad  \lim_{\lambda \rightarrow \infty} Z'  = 0,
 \label{ZLsubf}
\enq
the above result is similar to the case of a classical particle. Having calculated the partition function we are now in a position to calculate the probability density function. We present a two dimensional plot in figure \ref{fLsub2D},
and two cross section for low and high $\lambda$ in figures \ref{fLsubl} and \ref{fLsubh}.
It is clear that the Maxwell-Boltzmann approximation is only appropriate for low temperatures (high $\lambda$) but fails completely at high temperatures. Finally we calculate the average energy:
\ber
 \bar{E}'_{kLsub} (\lambda) &=& \frac{E[E_{kLsub}]}{m c^2} = E[\sqrt{1+p'^2}] =
4 \pi \int_{0}^{\infty}\sqrt{1+p'^2} f (\vec p') p'^2 dp'
\nonumber \\
&\hspace {-4 cm}=& \hspace {-2 cm} \frac{4 \pi}{Z' (\lambda)} \int_{0}^{\infty}\sqrt{1+p'^2} e^{-\lambda \sqrt{1+p'^2}} p'^2 dp' =
- \frac{1}{Z' (\lambda)} \frac{d Z' (\lambda) }{d\lambda}
= -  \frac{d \ln Z' (\lambda) }{d\lambda}
\label{avergaenergycfp2}
\enr
this expression can be evaluated numerically and is depicted in figure \ref{Eavh} for high
$\lambda$ and figure \ref{Eavl} for low $\lambda$. Again we notice that the classical approximation is only valid for high $\lambda$ (low temperature).
In any case (exact or approximated) the average energy is a decreasing function of $\lambda$ or an increasing function of temperature as might be expected. However, for Euclidean particles the results are less intuitive as will be shown below.

\subsubsection{An Euclidean particle}

\begin{figure}
\centering
\includegraphics[width= 0.7 \columnwidth]{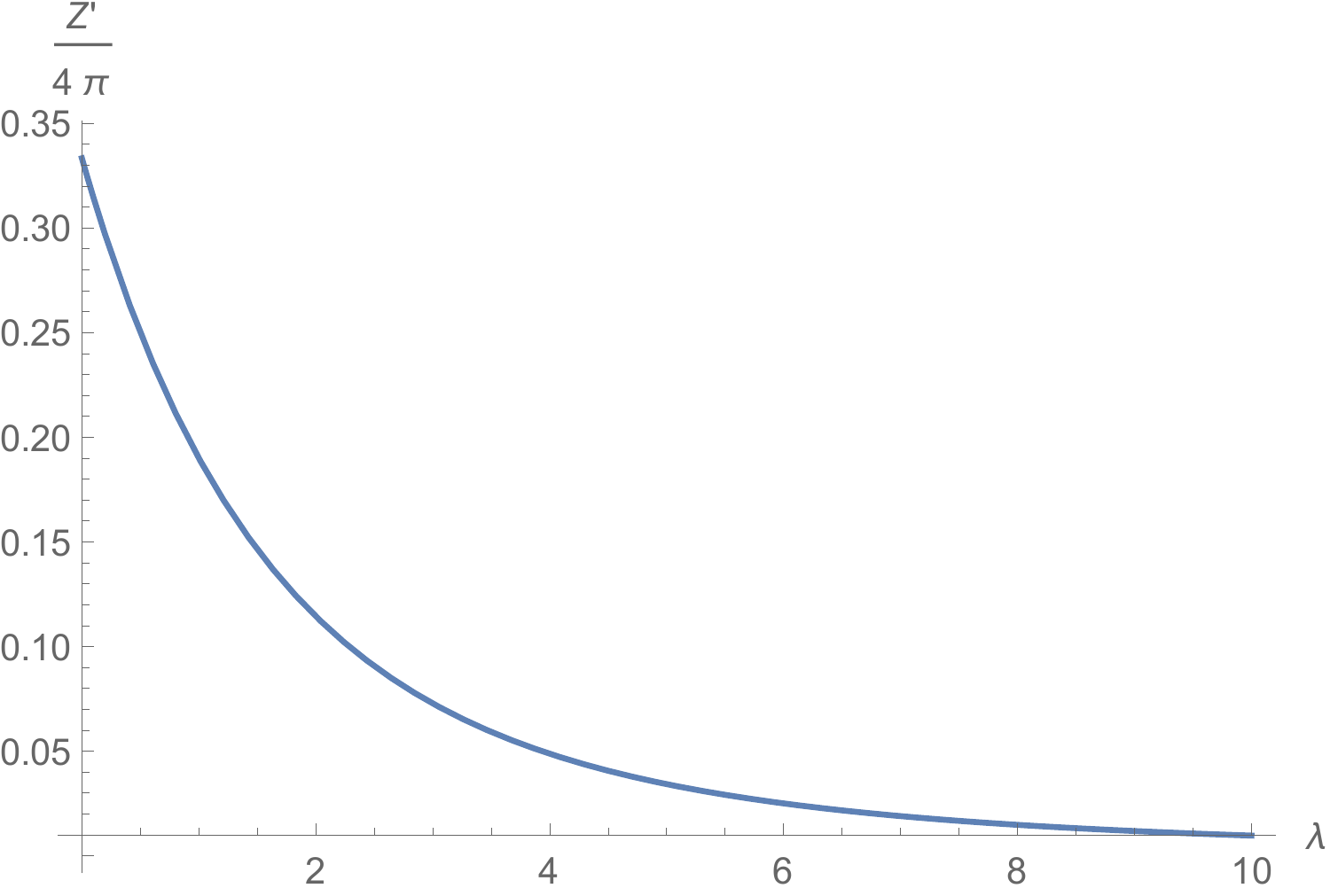}
 \caption{Partition function for a free Euclidean particle.}
 \label{ZEf}
\end{figure}

\begin{figure}
\centering
\includegraphics[width= 0.7 \columnwidth]{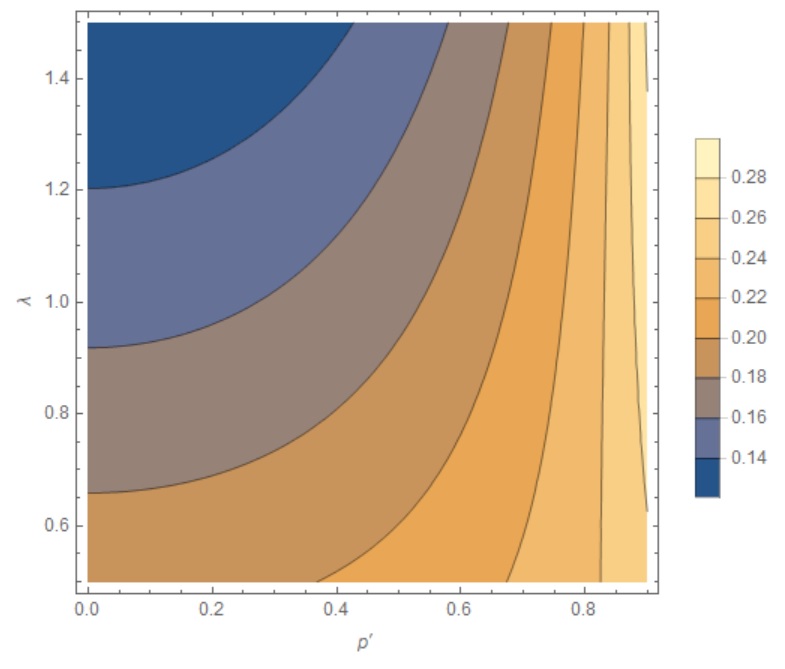}
 \caption{Probability density function for a free Euclidean particle as function of $p'$ and $\lambda$.}
 \label{fE2D}
\end{figure}

\begin{figure}
\centering
\includegraphics[width= 0.7 \columnwidth]{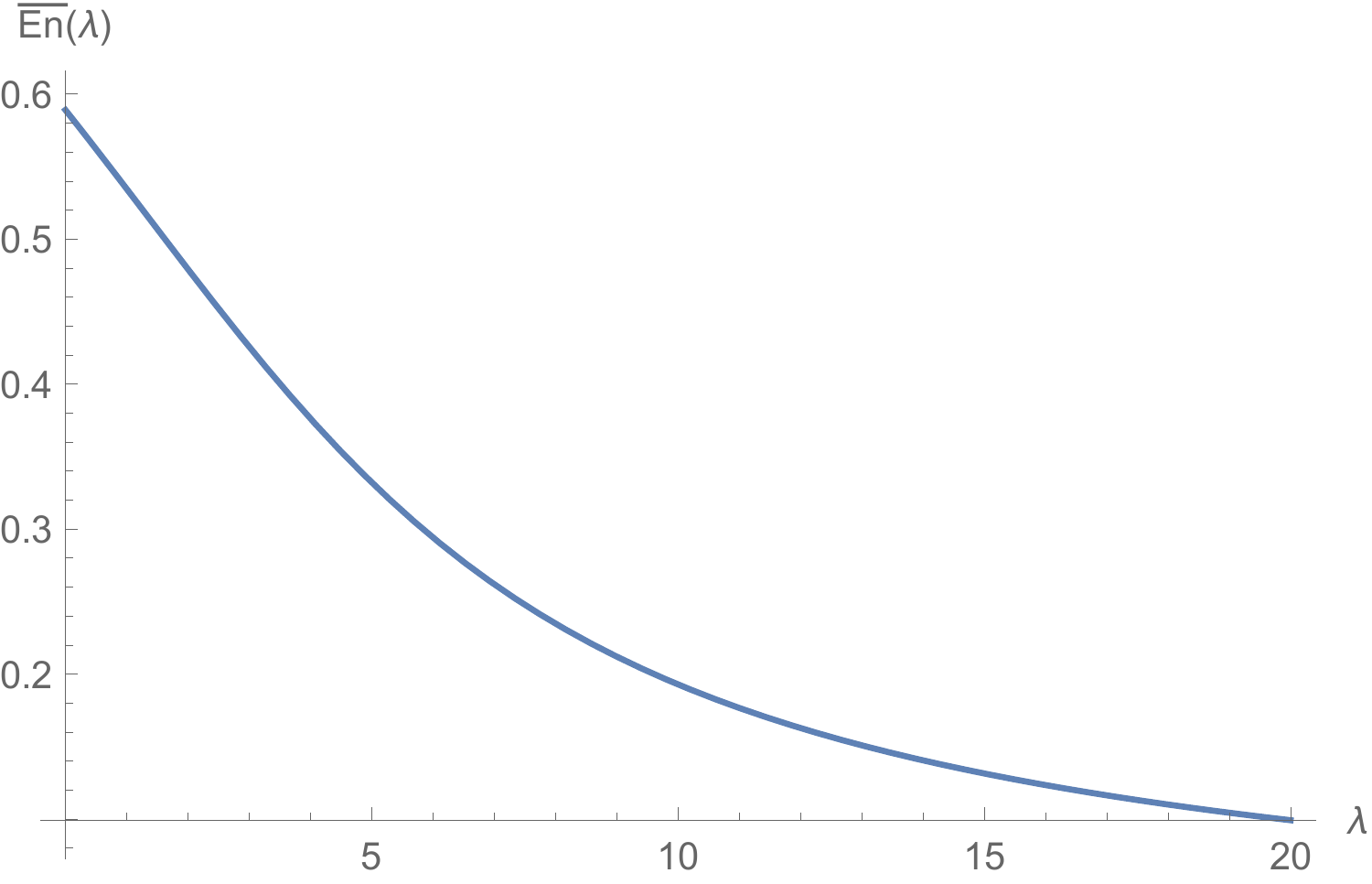}
 \caption{Average energy for free Euclidean particles with high $\lambda$.}
 \label{EavE}
\end{figure}

As we saw in previous sections, Euclidean particles share with Lorentzian particle the energy positivity property. However, they differ in the structure of their phase space considerably thus
taking into account \ern{Hp2} for a free particle, and \ern{fsys3} we arrive at:
\beq
 f (\vec p') = \frac{e^{-\beta_T H}}{Z'} = \frac{e^{-\lambda \sqrt{1-p'^2}}}{Z'}
\label{fE}
\enq
in which:
\beq
 Z' (\lambda) =  4 \pi \int_{0}^{1} e^{-\lambda \sqrt{1-p'^2}} p'^2 dp'
\label{ZE}
\enq
in which we recall that a free Euclidean particle phase space is compact and that $0 \le p' \le 1$.
The above expression cannot be evaluated analytically, but can be easily evaluated numerically
(see figure \ref{ZEf}). As can be clearly seen the partition function is a decreasing function of $\lambda$ or an increasing function of temperature.
For an Euclidean particle we have:
\beq
 \lim_{\lambda \rightarrow 0} Z'  = \frac{4 \pi}{3}, \qquad  \lim_{\lambda \rightarrow \infty} Z'  = 0,
 \label{ZE2}
\enq
Having calculated the partition function we are now in a position to calculate the probability density function. We present a two dimensional plot in figure \ref{fE2D},
and a cross section in figure \ref{fE}.
\begin{figure}
\centering
\includegraphics[width= 0.7 \columnwidth]{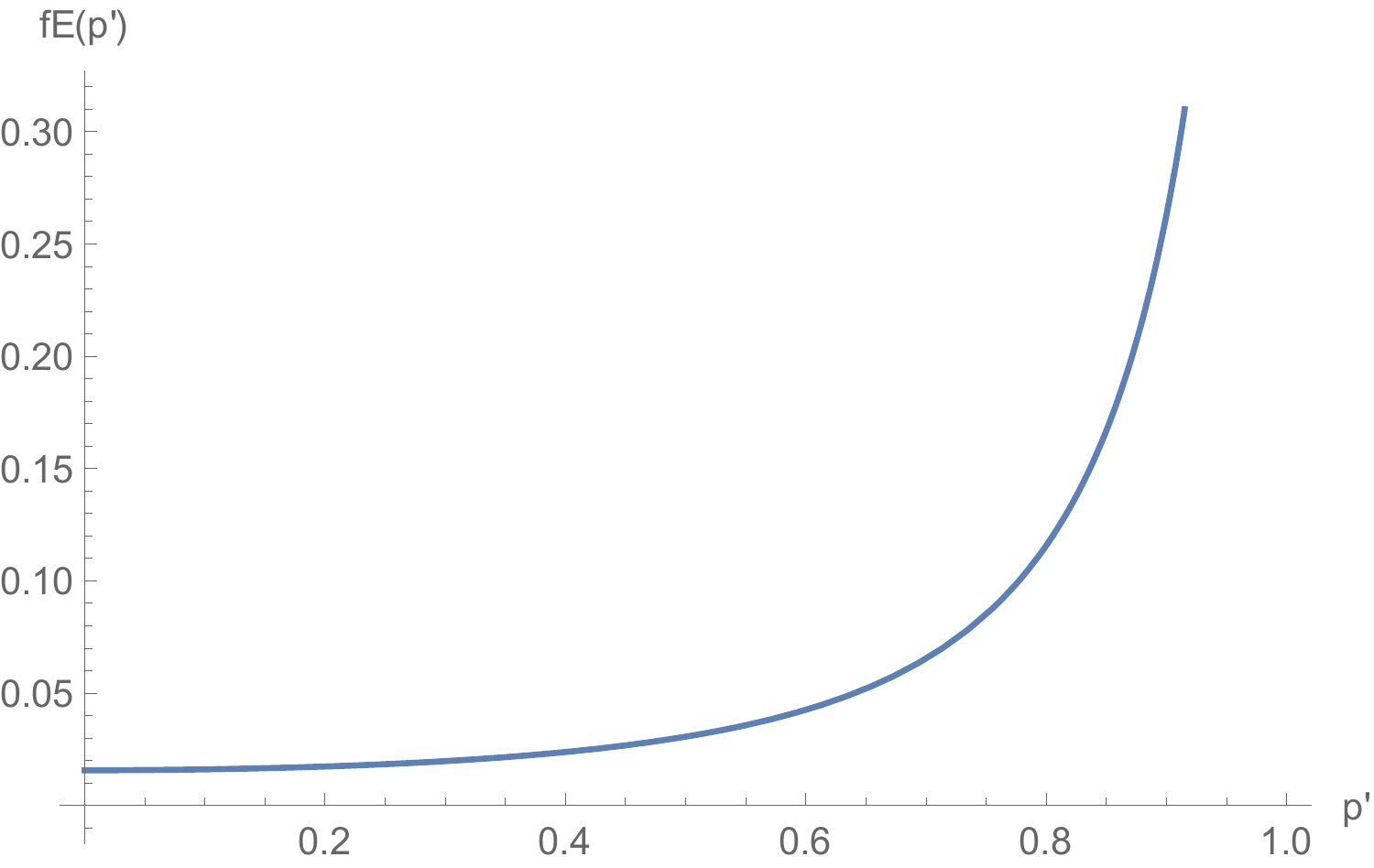}
 \caption{Probability density function for a free Euclidean particle as function of $p'$ for $\lambda = 5$.}
 \label{fE2}
\end{figure}
It is remarkable that it is more probable to find an Euclidean particle with high momentum than
in low momentum, this is in sharp contradiction to the situation for Lorentzian subluminal particles that prefer to stay in lower momenta. Of course in both cases high momenta means high velocity. However, this fact correlates well with the energy being a decreasing function of velocity in the Euclidean case. Finally we calculate the average energy:
\ber
 \bar{E}'_{kE} (\lambda) &=& \frac{E[E_{kE}]}{m c^2} = E[\sqrt{1-p'^2}] =
4 \pi \int_{0}^{1}\sqrt{1-p'^2} f (\vec p') p'^2 dp'
\nonumber \\
&\hspace {-4 cm}=& \hspace {-2 cm} \frac{4 \pi}{Z' (\lambda)} \int_{0}^{1}\sqrt{1-p'^2} e^{-\lambda \sqrt{1-p'^2}} p'^2 dp' =
- \frac{1}{Z' (\lambda)} \frac{d Z' (\lambda) }{d\lambda}
= -  \frac{d \ln Z' (\lambda) }{d\lambda}
\label{avergaenergyEfp}
\enr
this expression can be evaluated numerically and is depicted in figure \ref{EavE}.
Thus the average energy in the Euclidean case is a decreasing function of $\lambda$ as in the Lorentzian subluminal case. Thus it is an increasing function of temperature as might be expected. However, for Euclidean particles high temperature and high average energies entail low velocities. And low temperatures in the scale of $m c^2$ imply high velocities. A cooling down Euclidean universe will have a larger proportion of extremely fast moving particles.

\subsubsection{A Lorentzian superluminal particle}

\begin{figure}
\centering
\includegraphics[width= 0.7 \columnwidth]{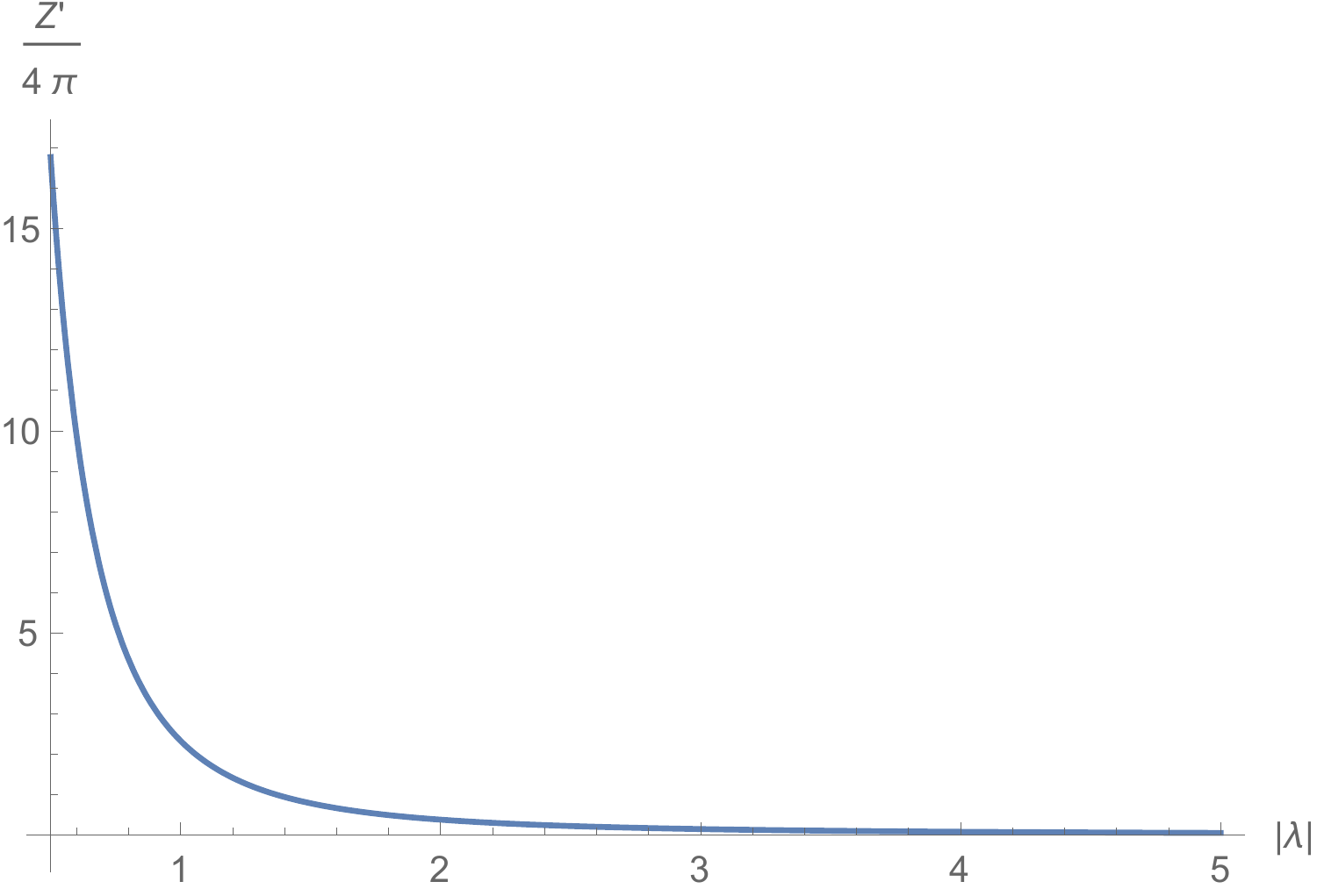}
 \caption{Partition function for a free superluminal Lorentzian particle.}
 \label{ZLsupf}
\end{figure}

\begin{figure}
\centering
\includegraphics[width= 0.7 \columnwidth]{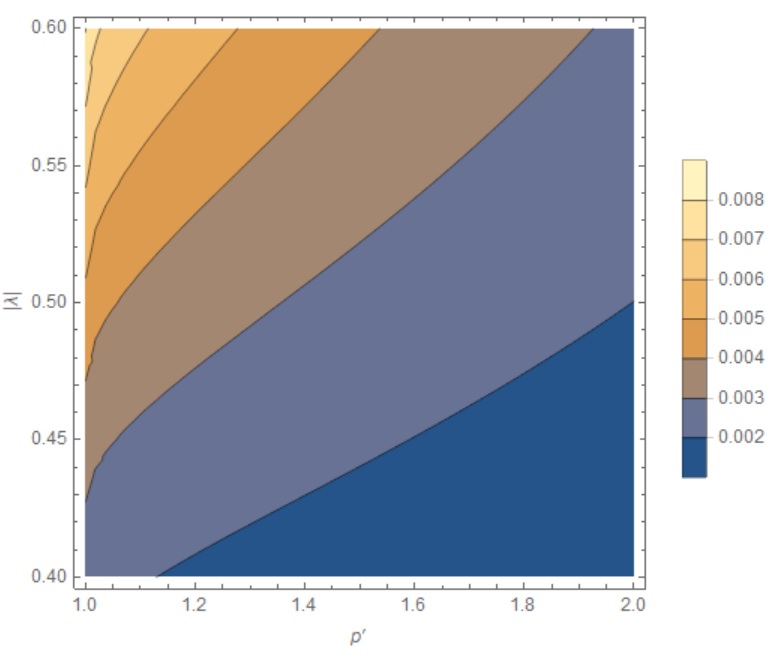}
 \caption{Probability density function for free superluminal Lorentzian particle as function of $p'$ and $|\lambda|$.}
 \label{fLsup2D}
\end{figure}

\begin{figure}
\centering
\includegraphics[width= 0.7 \columnwidth]{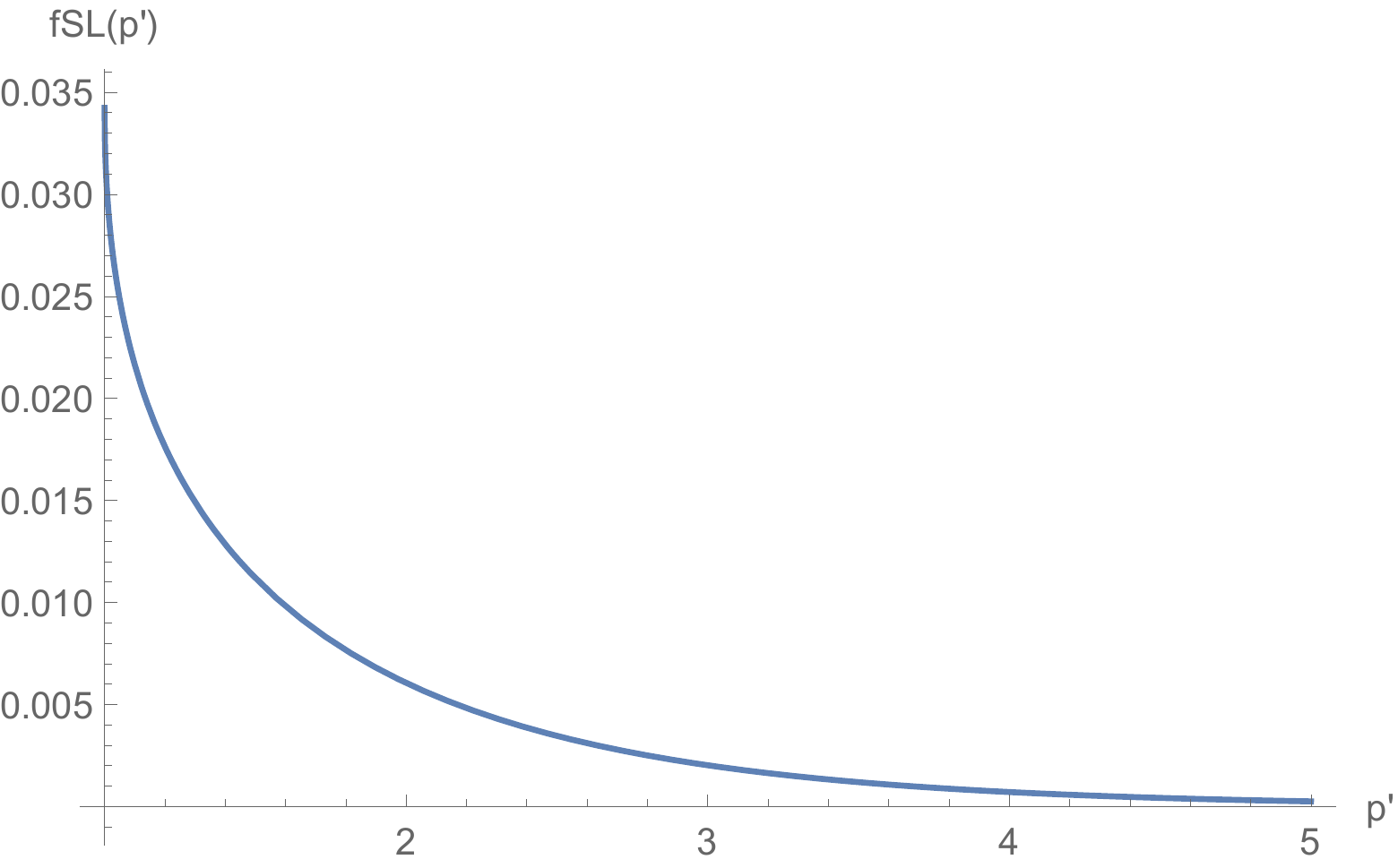}
 \caption{Probability density function for a free superluminal Lorentzian particle as function of $p'$ for $|\lambda| = 1$.}
 \label{fLsupf}
\end{figure}

\begin{figure}
\centering
\includegraphics[width= 0.7 \columnwidth]{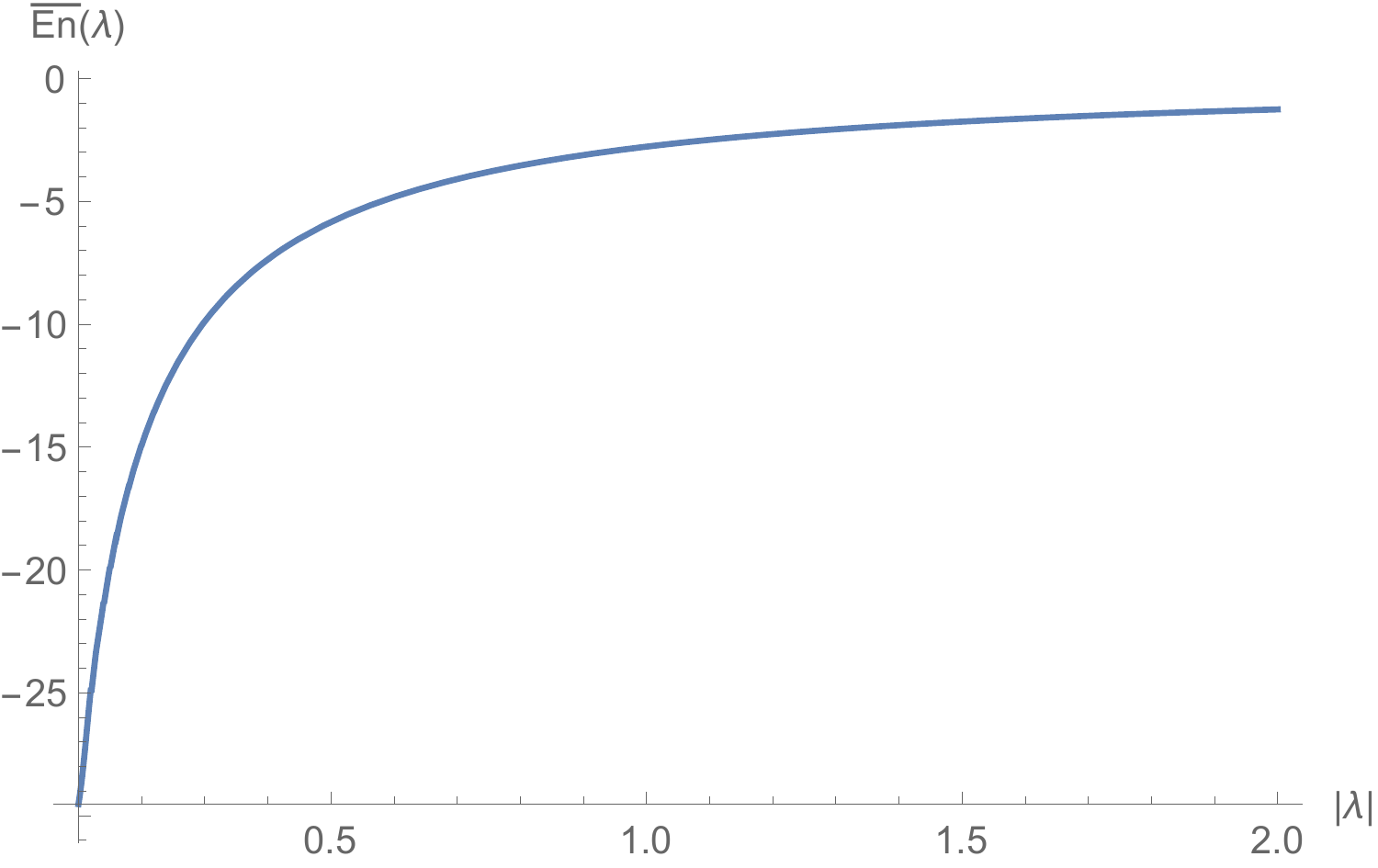}
 \caption{Average energy for free Lorentzian subluminal particles with high $|\lambda|$.}
 \label{EavLsup}
\end{figure}

As we saw in previous sections, the phase space of a free Lorentzian superluminal particle is complementary to the phase space of an Euclidean particle . Moreover, this case is unique with respect to the two previous cases as its free energy is negative. Taking into account \ern{Hp2} for a free particle, and \ern{fsys3} we arrive at:
\beq
 f (\vec p') = \frac{e^{-\beta_T H}}{Z'} = \frac{e^{\lambda \sqrt{p'^2-1}}}{Z'}
\label{fLsup}
\enq
in which:
\beq
 Z' (\lambda) =  4 \pi \int_{1}^{\infty} e^{\lambda \sqrt{p'^2-1}} p'^2 dp'
\label{ZLsup}
\enq
in which we recall that for a superluminal Lorentzian free particle phase space the unit sphere is excluded thus $1 < p'$. It easy to see that area below this probability density function
is infinite for every $\lambda>0$ and thus $Z'$ diverges. The proof is as a follows: choose any finite $p'_L >> 1$ thus we may write:
\beq
 Z' (\lambda) =  4 \pi \int_{1}^{p'_L} e^{\lambda \sqrt{p'^2-1}} p'^2 dp'
  + 4 \pi \int_{p'_L }^{\infty} e^{\lambda \sqrt{p'^2-1}} p'^2 dp'
\label{ZLsup2}
\enq
now:
 \beq
 4 \pi \int_{p'_L }^{\infty} e^{\lambda \sqrt{p'^2-1}} p'^2 dp' \simeq
 4 \pi \int_{p'_L }^{\infty} e^{\lambda p'} p'^2 dp'
\label{ZLsup3}
\enq
the right hand expression can be calculated analytically:
 \beq
  4 \pi \int_{p'_L }^{\infty} e^{\lambda p'} p'^2 dp'= e^{\lambda p'}
  \left.\left[\frac{p'^2}{\lambda} -  \frac{2p'}{\lambda^2}+  \frac{2}{\lambda^3}\right] \right|^\infty_{p'_L} = \infty
\label{ZLsup4}
\enq
hence $Z' (\lambda) = \infty$ for $\lambda>0$. There are two possible conclusions at this stage, either that a thermal equilibrium distribution is impossible for the superluminal Lorentzian free
particles, or that a thermal equilibrium does exist but with a negative $\lambda$ which entails a negative temperature. Admittedly this is a strange concept, however, if we are to accept superluminal Lorentzian particles in thermal equilibrium there is no way around it. Hence:
\beq
 f (\vec p') =  \frac{e^{-|\lambda| \sqrt{p'^2-1}}}{Z'}
\label{fLsup2}
\enq
in which:
\beq
 Z' (\lambda) =  4 \pi \int_{1}^{\infty} e^{-|\lambda| \sqrt{p'^2-1}} p'^2 dp'
\label{ZLsu5}
\enq
The above expression cannot be evaluated analytically, but can be easily evaluated numerically
(see figure \ref{ZLsupf}). As can be clearly seen the partition function is a decreasing function of $|\lambda|$ or an increasing function of temperature.
For a free superluminal Lorentzian particle we have:
\beq
 \lim_{|\lambda| \rightarrow 0} Z'  = \infty, \qquad  \lim_{|\lambda| \rightarrow \infty} Z'  = 0,
 \label{ZLsupf2}
\enq
Having calculated the partition function we are now in a position to calculate the probability density function. We present a two dimensional plot in figure \ref{fLsup2D},
and a cross section in figure \ref{fLsupf}.
Thus superluminal particles have a higher probability to be at lower momenta which is the situation for Lorentzian subluminal particles. However, for superluminal particles low momenta means high velocity (and not low velocity). Hence super luminal particles will tend to have $v \gg c$.
 Finally we calculate the average energy:
\ber
 \bar{E}'_{kE} (\lambda) &=& \frac{E[E_{kE}]}{m c^2} = - E[\sqrt{p'^2-1}] =
- 4 \pi \int_{1}^{\infty}\sqrt{p'^2-1} f (\vec p') p'^2 dp'
\nonumber \\
&\hspace {-4 cm}=& \hspace {-2 cm} -\frac{4 \pi}{Z' (\lambda)} \int_{1}^{\infty} \sqrt{p'^2-1} e^{-|\lambda| \sqrt{p'^2-1}} p'^2 dp' =
\frac{1}{Z' (\lambda)} \frac{d Z' (\lambda) }{d|\lambda|}
=  \frac{d \ln Z' (\lambda) }{d|\lambda|}
\label{avergaenergyLsupfp}
\enr
this expression can be evaluated numerically and is depicted in figure \ref{EavLsup}.
Thus the average energy or free Lorentzian superluminal particles is a increasing function of $|\lambda|$ contrary to the Lorentzian subluminal case. Thus it is an decreasing function of the absolute temperature but an increasing function of the true temperature as might be expected. We notice that for moderate temperatures the average energy flattens near zero energy level.

\section{Some Possible Cosmological and Physical Implications}

Suppose that the universe is Euclidean at $t=0$, once its starts to expand the temperature drops and the Euclidean particles become faster thus increasing the rate of the universe expansion and thermalization, obviously there is no horizon (homogeneity) problem for Euclidean particles. As the universe increases further the temperature continues to drop making the particles even faster, thus creating a positive feedback loop. This increased expansion is cosmological inflation, but without an ad hoc inflationary field \cite{Inflation}. This is the primordial particle accelerator of the cosmos. We notice that a Higgs type fields do not give the correct density perturbation spectrum \cite{Inflation}, hence one is forced to postulate a new field which is not a part of any particle model and thus is a possible but inelegant solution of the homogeneity problem. Alternatively
one can speculate that homogeneity is achieved by ordinary matter which can become superluminal as the current analysis shows.

However, as the universe expands to a certain limit the density drops and the Euclidean metric becomes unstable \cite{Yahaloma} and a Lorentzian metric develops instead. In a Lorentzian space-time we have two distinct particle species that cannot mix, the subluminal particles that we are familiar with, and the superluminal particles which tend to reach higher and higher velocities and are thus moving to the further reaches of the universe, quite beyond our reach. Those particles
may be what is perceived as dark energy \cite{Peebles} which affect the velocities of very distant supernovae and the CMB spectrum. $\Lambda CDM$ cosmology predicts that $0.76 \pm 0.02 \%$ of the universe are made of an unexplained "dark energy" component, obviously the Occam razor principle will vindicate a model in which such a ad-hoc component is not needed.

\begin{figure}
\centering
\includegraphics[width= 0.7 \columnwidth]{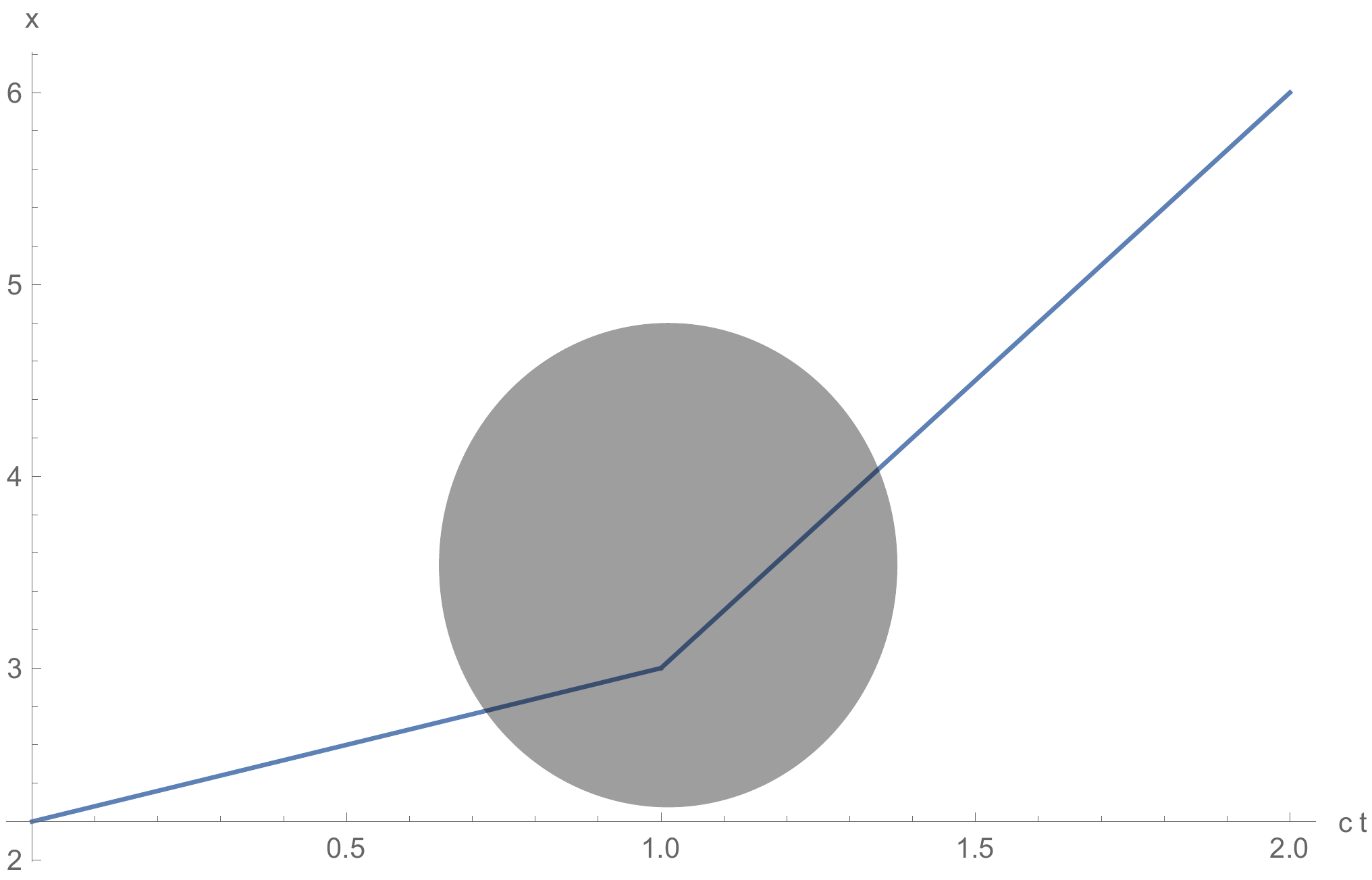}
 \caption{A schematic acceleration scheme of a particle in an Euclidean portion of space-time.}
 \label{accscheme}
\end{figure}

Another obvious physical implication of the previous analysis involve a far fetched technological scenario, is which a particle is accelerated to a velocity close to the velocity $c$ in a Lorentz space-time, enter into an artificially created Euclidean space-time and accelerated further in this region to velocities above the speed $c$ and emerge in a Lorentz space in which it will remain above the speed $c$ for ever unless it is decelerated in an Euclidean space again (see figure \ref{accscheme}).
This  may happen to a particle which travels radially in a Friedman-Lemaitre-Robertson-Walker metric passing outwards the critical radius of $r_c=\frac{1}{\sqrt{\kappa}}$ and then coming back at superluminal velocities. However, this will be very difficult to do artificially. Obviously a metric
change will require a significant $T_\mn$ according to \ern{ein}. Taking into account that the
largest metric deviation from the Lorentzian metric is the solar system on the surface of the sun in $h_{00} \sim10^{-6}$ \cite{Lensing}, it does not seem conceivable that such a metric change can be indeed implemented.

Last but not least one should remember that although classical physics is assumed to take place in a Lorentzian background, quantum field theory calculations are done in an Euclidean background using the Wick rotation. This is usually justified on the basis that it is  an analytic continuation. But an analytic continuation is a mathematical technique which has no physical justification
in Lorentzian space-time but makes perfect sense if part of space-time, in particular the part which is very close to the particle is Euclidean. Hence one may speculate that each elementary particle may carry with it a "bubble" of a microscopic Euclidean space-time.

\section{Conclusions}

We have shown that general relativity allows for non-Lorentzian space-times in particular this is allowed in part of the Friedman-Lemaitre-Robertson-Walker universe. The result of which is that superluminal particles can exist in such a cosmology. Some of the cosmological implications of superluminal particles regarding the homogeneity problem, and dark energy problems are underlined. Some other possible implications of non Lorentzian metrics which are not connected to superluminality but may be a consequence of non-Euclidean metrics are also suggested. Of course much more detailed analysis is needed to reach a definite conclusion regarding any of the above physical problems but the existence of non-Lorentzian space-times and superluminal particles suggests a plausible solution.

In the scope of the current paper we have only considered canonical ensembles in the number of particles is fixed, however, at high energies pair creation  from the vacuum is possible, hence a
grand canonical ensemble should be studied. Quantum mechanical effects were also out of the scope of the current paper which concentrated on classical effects only.

Finally, an Euclidean metric will effect the energy momentum tensor thus effecting the allowed solutions of the Friedman-Lemaitre-Robertson-Walker universe. An exact mathematical model describing
the transition from the Euclidean to the Lorentzian universe in which we leave in, is left for future studies.

\end{paracol}
\reftitle{References}




\end{document}